\documentstyle[12pt,aasms4]{article}

\lefthead{Kaiser et al.}
\righthead{NGC$\,$4151}

\begin{document}

\title{The Resolved Narrow Line Region in NGC$\,$4151\altaffilmark{1} }
\altaffiltext{1}{Based on observations with the NASA/ESA \it Hubble Space 
Telescope,
\rm obtained at the Space Telescope Science Institute, which is operated
by AURA Inc under NASA contract NAS5-26555}
\author{M. E. Kaiser\altaffilmark{2},
L.D. Bradley~II\altaffilmark{2},
J.B.Hutchings\altaffilmark{3},
D.M.Crenshaw\altaffilmark{4}, 
T.R.Gull\altaffilmark{5},
S.B.Kraemer\altaffilmark{4},
C. Nelson\altaffilmark{6},
J. Ruiz\altaffilmark{4},
D.Weistrop\altaffilmark{6}
}\
\altaffiltext{2}{Dept of Physics and Astronomy, The Johns Hopkins University, 
Baltimore, MD 21218}
\altaffiltext{3}{Dominion Astrophysical Observatory, National Research Council 
of Canada, 5071 W. Saanich Rd., Victoria B.C. V8X 4M6, Canada}
\altaffiltext{4}{Dept of Physics, Catholic University of America, Washington DC20064}
\altaffiltext{5}{NASA Goddard Space Flight Center, Lab for Astronomy and Solar
Physics, Code 681, Greenbelt MD 20771}
\altaffiltext{6}{Dept of Physics, University of Nevada, Las Vegas,
4505 Maryland Pkwy, Las Vegas, NV 89154-4002}

Subject headings: galaxies: Seyfert, galaxies: kinematics and dynamics

\begin{abstract}

We present slitless spectra of the Narrow Line Region (NLR) in
NGC$\,$4151 from the Space Telescope Imaging Spectrograph (STIS) on
{\it {HST}}, and investigate the kinematics and physical conditions of
the emission line clouds in this region.  Using medium resolution
($\sim\,$0.5$\,$\AA) slitless spectra at two roll angles and narrow
band undispersed images, we have mapped the NLR velocity field from
1.2$\,$kpc to within 13$\,$pc
(H$_{\circ}\,=\,$75$\,$km$\,$s$^{-1}$Mpc$^{-1}$) of the nucleus.  The
inner biconical cloud distribution exhibits recessional velocities
relative to the nucleus to the NE and approaching velocities to the SW
of the nucleus.  We find evidence for at least two kinematic
components in the NLR.  One kinematic component is characterized by
Low Velocities and Low Velocity Dispersions (LVLVD clouds:
$\mid$v$\mid\,<\,$400$\,$km$\,$s$^{-1}$, and
$\Delta$v$\,<\,$130$\,$km$\,$s$^{-1}$).  This population extends
through the NLR and their observed kinematics may be gravitationally
associated with the host galaxy.  Another component is characterized
by High Velocities and High Velocity Dispersions (HVHVD clouds:
400$\,<\,\mid$v$\mid\,{^{<}_{\sim}}\,$1700 km$\,$s$^{-1}$,
$\Delta$v$\,\geq\,$130$\,$km$\,$s$^{-1}$). This set of clouds is
located within 1.1$\arcsec$ ($\sim\,$70$\,$pc) of the nucleus and has
radial velocities which are too high to be gravitational in origin,
but show no strong correlation between velocity or velocity dispersion
and the position of the radio knots.  Outflow scenarios will be
discussed as the driving mechanism for these HVHVD clouds.  We also
find clouds characterized by Low Velocities and High Velocity
Dispersions (LVHVD clouds: $\mid$v$\mid\,<\,$400 km$\,$s$^{-1}$,
$\Delta$v$\,\geq\,$130$\,$km$\,$s$^{-1}$).  These clouds are located
within 3.2$\arcsec$ ($\sim\,$200$\,$pc) of the nucleus.  It is not
clear if the LVHVD clouds are HVHVD clouds whose low velocities are
the results of projection effects.

Within 3.2$\arcsec$ ($\sim\,$200$\,$pc) of the nucleus, the
[OIII]/H$\beta$ ratio declines roughly linearly for both the High
Velocity Dispersion (HVD) and LVLVD clouds. Since the ionization
parameter is proportional to r$^{-2}$n$^{-1}$, it appears that the
density, n, must decrease as $\sim$r$^{-1}$ for the clouds within the
inner $\sim\,$3.2$\arcsec$. At distances further from the nucleus, the
[OIII]/H$\beta$ ratio is roughly constant.

\end{abstract}

\section{Introduction}
\label{sec-intro}

Current models describing Seyfert galaxies consist of an active
nucleus surrounded by a dense
($\sim\,$10$^{9}\,-\,$10$^{10}$cm$^{-3}$) broad emission line region
(BLR) which lies well within $\sim\,$1 pc.  Extending to larger radii
($\sim\,$1$\,-\,$1000~pc) is a lower density
($^{<}_{\sim}\,$10$^{5}$cm$^{-3}$) region where the narrow permitted
and forbidden lines of the NLR are generated.  Nuclear activity is
believed to be generated by mass infall from a hot accretion disk
surrounding a massive black hole.  X-rays released by this process
irradiate the accretion disk generating a biconical wind.  This
nuclear wind may then be responsible for the outward acceleration of
clouds into the NLR.

Radio emitting jets have also been observed in Seyfert galaxies.  The
jet is believed to be aligned with the rotation axis of the accretion
disk and may provide another mechanism for cloud acceleration.  Line
profile studies of Seyferts with multi-knot radio jets have shown that
substructure in the [OIII] emission line profiles is correlated with
the spatial locations of radio jet knots, indicating that the
interaction with the radio plasma may produce density enhancements in
the NLR gas (\cite {whittle88}, \cite {aoki98}).  By sweeping up gas
the radio plasma may also serve to provide a path for nuclear photons,
thus facilitating ionization in the extended emission line region.

At distances large compared to the accretion disk the gravitational
influence of the host galaxy dominates the velocity field.  Gas in
this region is characterized by orbital rather than radial or
turbulent velocities.  

Although this complex model has emerged, it is far from verified in
every detail and the interaction between the NLR and its local
environment is still not well understood.  Emission line profiles show
abrupt spatial variations, thus supporting the theory that multiple
processes are influencing the NLR gas.  The blue asymmetry of the
narrow line profiles has led various authors to advocate either infall
(\cite {derobertis90}) or outflow (e.g.~\cite{schulz90}, \cite
{whittle92}). Line profiles have been modelled and matched to
observations for wind models which confine and drive the narrow line
clouds (\cite {krolik81}, \cite {krolik84}, \cite {schiano86a}, \cite
{schiano86b}).  Correlations have been found between substructure in
the emission line profiles and the position of the radio jets
(\cite{whittle88}, \cite{winge99}). However, distinguishing among the
various dynamic processes is inhibited by the low spatial resolution
or limited spatial mapping of the NLR.  Complete, high spatial
resolution mapping of the complex NLR gas distribution will be an
important contribution in distinguishing among the models.

Ground-based studies of the NLR have been limited to spatial
resolutions of ${^{>}_{\sim}}\,$50~pc for the closest Seyfert
galaxies.  In general they have relied upon long-slit spatial line
profile studies to understand the dynamics of the NLR and its
dependence upon the BLR (\cite {schulz90}, \cite
{veilleux91}). Fabry-Perot observations have mapped the entire
spatially extended NLR, but these observations are also ultimately
limited to coarse spatial resolution by seeing conditions
($\sim\,$60~pc resolution for NGC$\,$1068, \cite {cecil90}).

With the advent of {\it {HST}}, high spatial resolution imaging of
Seyfert galaxies in the optical has exposed a rich, complex and varied
NLR.  Narrow band WFPC images of [OIII] and H$\alpha\,+\,$[NII] have
revealed the morphological structure of the NLR for a large sample of
Seyfert galaxies (NGC$\,$4151: \cite {evans93}, \cite
{boksenberg95}). Long slit observations have provided kinematic
information at a few select positions within the NLR (e.g.,
NGC$\,$4151: \cite {winge97}, \cite {winge99}, Mrk$\,$3: \cite
{capetti98}).  While this complement of observations has expanded our
knowledge, the lack of full kinematic mapping still inhibits our
understanding of the environmental forces influencing the NLR.

As one of the brightest and closest
(v$_{recessional}\,\sim\,$997$\,$km$\,$s$^{-1}$, \cite {pedlar93})
Seyfert~1 galaxies, NGC$\,$4151 is an ideal candidate for studying the
full kinematic field of the NLR emission line gas.  Now, with STIS
slitless spectroscopy the entire dispersed NLR can be imaged with a
spectral resolution of 0.276$\,$\AA$\,$pix$^{-1}$ at [OIII] and a
spatial resolution of 0.051$\arcsec\,$pix$^{-1}$.

Our previous STIS slitless spectroscopy presented a kinematic map of
the NGC$\,$4151 NLR (\cite {hutchings98}). In this paper we present a
more complete kinematic map of the NGC$\,$4151 NLR at a second roll
angle with expanded wavelength coverage encompassing three wavelength
ranges ($\sim\,$3700$\,-\,$3800$\,$\AA,
$\sim\,$4800$\,-\,$5100$\,$\AA, $\sim\,$6300$\,-\,$6850$\,$\AA) to
include emission lines with a range of ionization potentials.
Expanding upon our earlier measurements (\cite {hutchings98}) we
present a table of velocities and velocity dispersions for 57 emission
line clouds.  [OIII] and H$\beta$ fluxes have been measured for 34 of
the clouds in this sample. Where possible [OII], [OI], and [SII]
fluxes have also been measured.  The velocities and fluxes have been
employed to search for radial structure in the kinematic and
ionization distributions.  Correlations with the spatial location of
the radio jet have been investigated.  Outflow, infall, and
gravitational models are also examined within the context of the
velocity, velocity dispersion, and flux distributions.

The observations are discussed in Section \ref{sec-observations}, the
image processing is described in Section \ref{sec-processing}, the
velocity measurements in Section \ref{sec-velocities}, and the flux
measurements in Section \ref{sec-flux}.  The results are presented in
Section 6 and discussed in Section 7.

\section{Observations}
\label{sec-observations}

Slitless spectra of the nuclear region of NGC$\,$4151 were obtained
using the Space Telescope Imaging Spectrograph (\cite{kimble98},
\cite{woodgate98}) aboard the Hubble Space Telescope.  These CCD
observations were acquired at a single epoch, with the dispersion
direction oriented -45$^{\circ}$ E of N. Three medium dispersion
grating tilts were selected to permit observation of the spectral
region encompassing the [OII] $\lambda$3727 emission
(Figure~\ref{blue}), the H$\beta$ and [OIII] $\lambda$5007 emission
(Figure~\ref{green}) and the [OI] $\lambda$6300, H$\alpha\,+$ [NII],
and [SII] emission structure (Figure~\ref{red}). Each format spans a
50$\arcsec$ field-of-view in the spatial direction.  The following
emission lines were also detected: H12 $\lambda$3751, [FeVII]
$\lambda$3760, H11 $\lambda$3771, H10 $\lambda$3798, H9 $\lambda$3835,
[NeIII] $\lambda$3869, H8$\,+\,$HeI $\lambda$3889, and H$\epsilon$
$\lambda$3970 $+$ [NeIII] $\lambda$3967. For these observations the
resolution is defined by the spatial extent of the emission feature.
Assuming a 2.0 pixel emission feature, which is equivalent to the
average instrument resolution for a point source, the resolution is
0.55$\,$\AA~ (45 km s$^{-1}$) at [OII] $\lambda$3727, 0.55$\,$\AA~ (33
km s$^{-1}$) at [OIII] $\lambda$5007, and $\sim\,$1.1$\,$\AA~ (51 km
s$^{-1}$) at H$\alpha$.  However, the near-nuclear emission line
clouds typically have velocity broadened widths which are then
convolved with the Optical Telescope Assembly (OTA) plus instrument
point spread function (PSF) to produce observed widths of
$^{>}_{\sim}$9 pixels FWHM, rendering it difficult to deconvolve the
H$\alpha\,+$[NII] emission line clouds.

To assist in identification of the dispersed cloud structure and to
facilitate measurement of the cloud velocities and velocity
dispersion, observations were also acquired with the STIS narrow band
[OII] filter and the WFPC2 narrow band filters: F502N, F658N, F390N,
F673N, and F656N with the Wide Field (WF) camera.  The F502N image was
used as the primary undispersed image.  Its FWHM bandpass encompasses
emission line clouds with velocities extending from
$-\,$1700$\,$km$\,$s$^{-1}$ through $\sim\,$500$\,$km$\,$s$^{-1}$ with
respect to the systemic velocity.  The broadest wavelength coverage
was provided by the STIS [OII] filter whose bandpass includes
velocities extending from $-\,$2700$\,$km$\,$s$^{-1}$ through
2350$\,$km$\,$s$^{-1}$ with respect to the nucleus. The STIS [OII]
filter was used to confirm cloud identifications and the STIS
dispersed images at PA$\,=\,$104$^{\circ}$ and PA$\,=\,$45$^{\circ}$
were used to search for high velocity clouds not contained in the
F502N [OIII] bandpass.  With the exception of the F502N [OIII] image,
the WF narrow-band filter observations were acquired at a different
epoch than the dispersed images. Consequently, the position angle for
the WF images is at -41$^{\circ}$ E of N, whereas the position angle
is 45$^{\circ}$ E of N for the dispersed images and the WF [OIII]
image.

Previous STIS data (\cite{hutchings98}) were acquired using the same
instrument configuration as the current [OIII] dispersed data.  The
data acquired at this earlier epoch were oriented with the dispersion
direction 14$^{\circ}$ E of N. A log of the observations is presented
in Table 1; entries marked with an $^{a}$ represent the earlier
data. The archival WFPC2 data were acquired with the Planetary Camera
(PC) channel. The higher resolution of the PC compared to the WF
camera (0.0455 $\arcsec$/pixel versus 0.996 $\arcsec$/pixel) is
advantageous for identifying the numerous emission line clouds in the
near-nuclear region. These clouds are annotated in Figures~\ref{o3b},
\ref{o3a}, and \ref{o3d}.

\section{Image processing}
\label{sec-processing}

Minimal post-pipeline processing was required for the WF and PC images
used to determine the spatial location of the [OIII] emission line
clouds relative to the NGC$\,$4151 nucleus.  This additional
processing consisted of rescaling the images to the STIS spatial scale
(0.05071 $\arcsec$/pixel) and then rotating the WF and PC images to
the same orientation as the STIS image.

For the STIS slitless spectra, scattered continuum light and the host
galaxy background were subtracted from the post-pipeline processed
images.  This was done to enhance the contrast of low surface
brightness clouds for the velocity measurements and to reduce the
background for the flux measurements.  Our background subtracted image
exhibits no evidence for uncorrected absorption of H$\alpha$ near the
nucleus.

The background template was constructed by median filtering the image
along the dispersion direction.  This template fit the background well
at distances greater than 1.6$\arcsec$ from the nucleus.  However, the
inner wings and the core of this spatial template varied along the
dispersion direction, due at least in part, to the variation of the
continuum as a function of wavelength and the scattered light from the
nucleus and near-nuclear clouds in the emission lines.  Consequently,
multiple templates were constructed as a function of wavelength to
subtract the near-nuclear region. Far-field template construction
consists of masking the emission line regions, and median filtering
the image along the dispersion direction.  The near-nuclear region
($\pm$1.6$\arcsec$) is then masked and the far-field template
subtracted from the entire image. The near-nuclear templates are
similarly constructed, using rows within $\pm$1.6$\arcsec$ of the
nucleus, but typically only 30 columns are averaged to preserve the
spectral signature of the continuum.  For the emission line regions,
the near-nuclear background template is an average of the neighboring
templates redward and blueward of each emission line region.

\section{Velocity Measurements}
\label{sec-velocities}

The [OIII] $\lambda$5007 emission is bright and unblended with other
emission lines, and therefore the [OIII] dispersed image and the F502N
PC images were selected as the primary images for determining the
cloud positions and dispersions. Since the NLR presents a well-ordered
velocity field, associating the dispersed clouds with the undispersed
counterparts is not difficult in general. In congested regions where
these identifications are more difficult, measurements of the spatial
locations of the clouds in the WFPC2 and STIS images by various
co-authors were required to agree within 1 pixel.

Multiple approaches were employed to measure the velocity displacement
and dispersion of the emission line clouds.  Measurements were made
using both background subtracted and non-background subtracted images,
using automated and purely interactive cloud identifications and fits.
Checks were also performed using the STIS [OII] undispersed image with
the H$\beta$ and [OI] dispersed images.  The errors quoted in Table 2
are the one sigma spread in the measurements obtained by multiple
co-authors using multiple techniques.  The quoted errors dominate the
statistical errors associated with Gaussian fits to the line profiles.
The narrow band filter data were inspected to check for ghosts and the
presence of the clouds in the various emission lines.

All cloud velocities were determined relative to the NGC$\,$4151
nucleus.  The location of the nucleus was determined by fitting a
Gaussian to the continuum of the non-background subtracted image in
both the spatial and dispersion directions.  Positions and velocity
dispersions for the individual clouds were determined by fitting
Gaussians to the extended spectral region encompassing the clouds.  In
general, the extent and proximity of the clouds to one another
required fitting multiple Gaussians with a linear background to an
extended region neighboring the cloud.  With this method, the cloud
centers with respect to the nucleus and the cloud FWHM are determined
for both the dispersed and undispersed images.  Both the relative
position and FWHM of each cloud were then corrected for anamorphic
magnification.  This correction demagnifies the STIS pixel size in the
dispersion direction by 11\%.  The velocity dispersion is then
determined by subtracting the instrument profile convolved with the
spatial extent of the cloud, defined as the FWHM of the undispersed
cloud, in quadrature from the dispersed FWHM of the cloud.

At large distances perpendicular to the dispersion direction, a small
misalignment in the rotation of the STIS and PC images translates into
an error in the cloud velocity.  At the location of cloud 44, which
lies 10.9$\arcsec$ from the nucleus, a 0.3$^{\circ}$ error in the
alignment of the STIS and PC images corresponds to a velocity shift of
20$\,$km$\,$s$^{-1}$ in the [OIII] emission.  Consequently, the outer
NLR cloud velocities are small and potentially consistent with zero.

In addition to the manual method, an automated procedure was written
for matching the dispersed emission line features with their
counterparts in the narrow-band images and calculating the radial
velocities and velocity dispersions for the individual clouds. With
this method, the images are aligned and registered.  Then ``slices''
parallel to the dispersion axis, at the same spatial offset from the
zero-point, are plotted for each image. Using a graphics cursor,
features are marked and labeled interactively such that peaks in the
dispersed image slice are matched with peaks in the undispersed image
slice.  Widths and centers of profile cuts at 80\%, 50\%, and 20\% of
the peak height are evaluated, similar to the standard techniques used
in analysis of emission line profiles (e.g. \cite{whittle85}). The
relative shift between the centers obtained in any of these ways is
converted into a shift in wavelength and then a velocity.  The
velocity dispersion is determined by subtracting the width of the
undispersed image in quadrature from the width in the dispersed image,
converting to a width in angstroms and then to a velocity width.

The aforementioned methods were suitable for associating the majority
of the dispersed clouds with their undispersed counterparts.  However,
given the cloud congestion in selected near-nuclear regions and the
potential for spatial-spectral ambiguity in the dispersion direction
with slitless spectra, a third method was employed. This method was
used to confirm cloud identification/positions in the dispersed image
(i.e. cloud regions 7-10-12 and 19-22-23), as well as identifying
cloud velocities and positions for those clouds which were either too
faint to be seen in the PC image or redshifted out of the PC F502N
filter bandpass (i.e. clouds with velocities $^{>}_{\sim}\,$500
km$\,$s$^{-1}$).  By employing dispersed images at two different
position angles the velocity and rest position of clouds with respect
to the nucleus can be determined.  Figure~\ref{roll} illustrates the
method.  Here we see that the second image has been rotated to the
position angle of the first image and the images have been aligned at
the nucleus.  The same cloud is identified in each image (i.e. cloud
A1, A2). To determine the undispersed cloud position, a line is
extended from the cloud parallel to the continuum in each image.  The
intersection of these two lines is the location of the undispersed
cloud (i.e. point A).  The distance from the undispersed to the
dispersed position should be the same in both images, i.e. A-A1 and
A-A2 should be equal.  Requiring that the velocity shift be the same
in both images is a powerful tool for resolving cloud identification
ambiguities in congested regions.

Since several of these clouds are faint, a calculation was used to
verify the cloud identifications determined from the geometric method
described above.  For the analytic method, the position
(x$_{mr}$,y$_{mr}$) of the cloud with respect to the nucleus in the
archival (unrotated) image was measured, and then the ansatz
y-position (y$^{\prime}_{\circ}$) of the cloud in the rotated image
was inserted into the combined equations 1 and 2 to determine the
x-position of the cloud in the rotated image, (x$^{\prime}_{\circ} +
\Delta$). Then the velocity and spatial position of the cloud in the
rotated cycle 7 image were calculated.

\begin{equation}
(x^{\prime}_{\circ} + \Delta) = x_{mr} + \Delta(1-cos(\theta))
\end{equation}

\begin{equation}
y^{\prime}_{\circ} =y_{mr} + \Delta sin(\theta)
\end{equation}

\begin{equation}
x_{mr} = (x_{\circ} + \Delta)cos(\theta) + y_{\circ}sin(\theta)
\end{equation}

\begin{equation}
y_{mr} = - (x_{\circ} + \Delta)sin(\theta) +y_{\circ}cos(\theta)
\end{equation}

In the rotated cycle 7 frame the spatial coordinates of the cloud with
respect to the nucleus are (x$^{\prime}_{\circ}$,
y$^{\prime}_{\circ}$), in the unrotated archival frame the coordinates
with respect to the nucleus are (x$_{\circ}$,y$_{\circ}$), $\theta$ is
the angle between the position angles of the two images, and $\Delta$
is the x-distance from the spatial location to the dispersed location
of the cloud. Clouds whose velocities were confirmed analytically are
presented in Table 3.

A velocity dispersion was calculated for each cloud.  The
instrument-convolved spatial width ($\Delta\,v_{i,PC}$) was determined
by a Gaussian fit to the cloud in the F502N PC image.  This width was
subtracted in quadrature from the Gaussian width fit to the cloud in
the dispersed STIS image ($\Delta\,v_{i,STIS}$).  The velocity
dispersions ($\Delta\,v_{i}\, = \,\sqrt{\Delta\,v_{i,STIS}^2 -
\Delta\,v_{i,PC}^2}\,$) are presented in
Table~\label{tbl-velocities}2.  Systematic errors dominate the
statistical errors.  The errors quoted in the table are the
1$\,\sigma$ spread of the measurements by the various authors.

For the high velocity clouds presented in Table~\label{tbl-highv}3,
there are no corresponding clouds identified in the PC image because
these are relatively faint clouds which have been Doppler shifted to
to the edge or outside the F502N filter bandpass.  For these clouds,
the velocity dispersion was calculated by subtracting in quadrature
the average instrument convolved spatial width ($<\Delta\,v_{PC}>_i$)
for a set of clouds who have approximately the same value for
$\Delta\,v_{i,STIS}$ as the high velocity cloud.

In order to determine whether the high velocity, high velocity
dispersion clouds might simply be the superposition of several low
velocity dispersion clouds, we performed a Monte Carlo simulation. An
ensemble of 3 or 4 clouds with a Gaussian FWHM of a typical LVD cloud
were superposed with fluxes ranging over a factor of 4 and velocities
separations of up to 300$\,$km$\,$s$^{-1}$. The observed distribution
of clouds on the sky makes this a reasonable number to superpose,
since by symmetry a cylinder of typical cloud diameter, oriented
perpendicular to the bicone axis, encompasses 4 clouds.  In addition,
fluxes fainter than about 1/4 of the brightest in a group have no
visible effect on the combined profile.  In over 95\% of the cases,
the resultant blends are very structured or multiple-peaked. The
observed high velocity dispersion profiles are all smooth. Thus, we
conclude that the broader line clouds do indeed have intrinsic high
velocity dispersions.

\section{Flux Measurements}
\label{sec-flux}

H$\beta$ and [OIII] fluxes have been measured for 34 of the 57
tabulated clouds.  In addition, we have measured the [OII], [OI], and
[SII] flux for a subset of bright, relatively isolated clouds.

Flux measurements were executed using the background subtracted
images.  Initially, clouds were fit with a linear background and the
minimum number of Gaussians, in the dispersion direction, required to
obtain a reasonable fit to the cloud and its neighbors.  In general,
neighboring clouds were fit simultaneously with the target cloud
because the wings of adjacent clouds overlap.  Fits were performed in
sequential rows spanning the flux peak in the spatial direction until
the cloud was indistinguishable from the background.  The total flux
for each cloud was determined by summing the area associated with the
Gaussian fit to each row encompassed by the cloud.  Typically clouds
spanned $\sim$5$-$7 rows ($\sim$0.25 - 0.35$\arcsec$) in the spatial
direction.  Flux ratios are calculated using the same number of rows
for the flux of each species.  The fluxes listed in Table~5 for
H$\beta$ and [OIII] may have lower values than the fluxes listed in
Table~4 because fewer rows were summed to obtain the [OII], [OI], and
[SII] fluxes, and hence the [OIII] and H$\beta$ fluxes and flux ratios
presented in Table~5.

Gaussian fits to the clouds tend to underestimate the wings of the
line profiles.  A sub-sample of relatively isolated clouds with
sufficient signal-to-noise in the [OII] image were modelled using both
Lorentzian and Gaussian line profiles. In general, the Lorentzian
provides a better fit to the wings.  However, the line profile wings
are probably elevated by scattered light within the emission line
region.  Due to the complexity of the emission structure it is
difficult to determine this background.  Consequently, the difference
between fluxes obtained with the Gaussian and Lorentzian line profile
models is treated as a measure of our systematic error.  The errors
quoted in Table~5 are the difference between the fluxes calculated
from the Lorentzian and Gaussian fits to the data.  These errors are
positive indicating that the Lorentzian fits yield higher fluxes, as
expected.

Fits to the [OII] doublet $\lambda\lambda$3729,3726 were impeded by
the lower flux of the [OII] emission and the cloud congestion in the
narrow line region (NLR) which was exacerbated by the 10 pixel
(2.8$\,$\AA) separation of the doublet.  Several strategies were
employed to maximize the number of [OII] clouds fit while maintaining
relatively low errors on the measured flux.  The best fits were
obtained by constraining the widths of the doublet components to be
equal while permitting the relative amplitude of the doublet
components and the velocity width of the cloud itself to vary.

The impact of fitting single rows, which subsample the data, was
assessed by summing 3 rows in the spatial direction and then
performing the fit.  These fits were consistent to within 2\% for the
sum of the [OII] components.

We were unable to correct these measurements for extinction because
the H$\alpha\,+\,$[NII] emission line clouds are too confused to
extract unblended H$\alpha$ flux measurements for comparison with the
H$\beta$ fluxes.  Our long slit data will reduce the scattered light
contribution and provide more extended spectral coverage using the
HeII lines at 1640$\,$\AA~and 4686$\,$\AA~for extinction correction.

\section{Results}
\label{sec-results}

{\it {Morphology:}} The NLR emission line clouds of NGC$\,$4151
exhibit a biconical distribution. This is evident in our STIS [OII]
image and the WFPC2 narrow band image (Figure~\ref{o3b}) and has been
illustrated previously with high spatial resolution {\it {HST}} images
(\cite{evans93}, \cite{boksenberg95}, \cite{winge97},
\cite{hutchings98}).  In both the dispersed and undispersed images, we
find that clouds to the SW of the nucleus are brighter, more numerous,
and have a more extended distribution than the clouds to the NE. These
results are consistent with both ground based (e.g.~\cite {penston90})
and previous {\it {HST}} measurements.  For [OIII] the emission line
gas in the slitless spectrum is detected $\sim\,$1.3$\,$kpc SW of the
nucleus and $\sim\,$500$\,$pc to the NE.  The [OII], H$\beta$, and
H$\alpha\,+\,$[NII] emission is less extended than this, spanning
$\sim\,$700$\,$pc SW of the nucleus and $\sim\,$300$\,$pc to the NE
(Figures~\ref{blue} and \ref{red}).  In addition, we detect [OI] and
[SII] emission line clouds in the NLR with a similarly asymmetric
spatial distribution.  Background subtraction of the STIS dispersed
[OII] image (Figure~\ref{blue}) also reveals NLR emission structure of
[FeVII] $\lambda$3760, [NeIII] $\lambda$3869, HeI $\lambda$3889, and
[NeIII] $\lambda$3967, as well as emission from higher level
transitions in the Balmer series (H$\epsilon$ $\lambda$3970 through
H12 $\lambda$3751).

{\it {Velocities and Dispersions:}} Our velocity and velocity
dispersion measurements of 57 emission line clouds
(Tables~\label{tbl-velocities}2 and \label{tbl-highv}3;
Figures~\ref{rv}, \ref{width}, and \ref{hist}) suggest that there are
at least two distinct kinematic components present in the NLR of
NGC$\,$4151. One component (LVLVD) is characterized by low velocities
and low velocity dispersions ($\mid$v$\mid\,<\,$400$\,$km$\,$s$^{-1}$
and $\Delta$v$\,<\,$130$\,$km$\,$s$^{-1}$).  A second cloud population
(HVHVD) is characterized by high velocities and velocity dispersions
($\mid$v$\mid\,>\,$400$\,$km$\,$s$^{-1}$ and
$\Delta$v$\,\geq\,$130$\,$km$\,$s$^{-1}$).  An apparent intermediate
cloud population with low velocities and high velocity dispersions,
the LVHVD clouds, may be HVHVD clouds whose low velocities are the
results of projection effects.

The HVD clouds all reside within 3.2$\arcsec$ ($\sim\,$200$\,$pc) of
the nucleus. The HVHVD clouds, which reside within 1.1$\arcsec$
($\sim\,$70$\,$pc), are more tightly clustered about the nucleus than
the LVHVD clouds. The LVLVD cloud population extends from the nucleus
out to a distance of $\sim\,$19$\arcsec$ (1.2$\,$kpc), with a gap in
the cloud distribution at a distance of $\sim$4.0$\arcsec$.  We may
also subdivide the LVLVD population into clouds clustered near the
nucleus and those beyond $\sim$4.0$\arcsec$ whose dynamics are
probably unaffected by nuclear processes.

{\it {Jet-cloud interactions:}} The high dispersion and highest
velocity clouds tend to be located near the radio jet axis
(Figure~\ref{radio}).  However, none of the HVHVD clouds are
coincident with the radio jet substructure.

The NGC$\,$4151 radio jet contains five distinct knots
(Figure~\ref{radio_map}, nomenclature follows \cite{carral90}) which
are not co-aligned along a single position angle, but can be
represented by an average position angle of 77$^{\circ}$
(\cite{pedlar93}). Higher resolution measurements indicate that radio
knot C4 consists of two components separated by 92$\:$milliarcseconds
(VLBI: \cite {harrison86} and VLBA: \cite {ulvestad98}).  The
brightest knot within the eastern component is co-spatial with the
optical nucleus (\cite {ulvestad98}). Since the radio data presented
in Figure~\ref{radio_map} (\cite {mundell95}) has a resolution of
0.16$\,\times\,$0.15$\arcsec$, we aligned the center of the C4 radio
component with the optical nucleus.  To align the two nuclei it was
necessary to translate the optical image by 0.06$\arcsec$ in right
ascension and 0.46$\arcsec$ in declination. Radio knot C3, at
$\sim\,$0.5$\arcsec$ SW of the nucleus, is only 0.19$\arcsec$ from the
projected position of cloud 20.  These two inner knots, C4 and C3, lie
along a position angle of 83$^{\circ}$.  Outer radio knots C2 and C5,
at $\sim\,$0.9$\arcsec$ SW and NE of the nucleus, lie along a position
angle of $\sim\,$73$^{\circ}$ (Figure~\ref{radio}).  C1, at
$\sim\,$1.9$\arcsec$ SW of the nucleus, is relatively diffuse and
isolated from the [OIII] emission line clouds.

The NLR [OIII] gas within $\sim\,$1.5$\arcsec$ of the nucleus, a
point defined by the 5$\,\sigma$ radio contour associated with radio
knots C2 and C5, is approximately co-spatial with the radio emission
along PA$\,=\,$77$^{\circ}$. We find that all the clouds in this
region are contained within an opening angle of $\sim\,$45$^{\circ}$
about the NLR axis at 63$^{\circ}$ with the HVD clouds relatively
symmetrically distributed about this axis.  We have overlaid
5$\,\sigma$ radio contours on the optical image
(Figure~\ref{radio_map}).  Of the 13 clouds contained within the
5$\,\sigma$ (1.7$\,$mJy) radio contour, 4 are LVD clouds.  Relaxing
the radio contour to 3$\,\sigma$ (0.96$\,$mJy), only increases the
cloud population by 2 HVD clouds. So although the [OIII] emission is
approximately aligned with the 3$\,\sigma$ and 5$\,\sigma$ radio
contour, the clouds interior to the these contours are both LVD and
HVD clouds.  Within a 1.5$\arcsec$ radius of the nucleus but exterior
to the 3$\,\sigma$ radio contour, we find a similar distribution of
LVD to HVD clouds.  Of the 10 clouds in this region, only 20\% are LVD
clouds.  Thus the radio jet does not appear to be responsible for the
HVD clouds.  In addition, none of the HVHVD clouds are coincident with
the radio axis and only 5 of 9 HVHVD clouds lie within the 3$\,\sigma$
radio contour. 

Cloud 20, whose projected position lies near radio knot C3, is a LVLVD
cloud which exhibits no kinematic or morphological signatures of
interaction with the radio jet.

{\it {Flux Distribution:}} Flux ratio measurements from different
lines in individual clouds are consistent with photoionization.

The radial distribution of H$\beta$ and [OIII] fluxes
(Figure~\ref{fluxes}) indicates that the flux upper envelope decreases
as a function of radius for r$\,<\,$3.2$\arcsec$ ($\sim\,$200$\,$pc).
At larger radii the flux is approximately constant.

Within a $\sim\,$200$\,$pc radius of the nucleus, the [OIII]/H$\beta$
flux ratio declines linearly for both the redshifted and blueshifted
clouds.  This behavior, albeit more pronounced for the HVD clouds, is
exhibited for both the HVD and LVD clouds (Figure~\ref{fluxes}).  This
decrease in the flux ratio indicates a decreasing ionization parameter
(\cite{ferland83}). At larger radii the flux ratio is comparable to
its value very close to the nucleus.

To further illustrate the spatial structure in the [OIII]/H$\beta$
ratio we have generated a [OIII]/H$\beta$ image (Figure~\ref{ratio}).
Much of the discrete cloud structure evident in the dispersed [OIII]
(Figure~\ref{green}) and H$\beta$ images is preserved in the ratio
image.  

Due to the small number of clouds for which we have [OII] fluxes, we
were not able to determine if the [OII]/H$\beta$ ratio exhibits the
same radial gradient in the inner NLR as presented by the high
velocity dispersion clouds for [OIII]/H$\beta$.

{\it {Continuum NLR Emission:}} As a further probe of the impact of
the radio emission on the observed optical emission structure, we have
searched the continuum emission associated with the optical clouds
near the radio emitting cores for an elevated blue continuum flux.
This excess flux at shorter wavelengths is a signature of free-free
emission and scattered light.  The
continuum flux redward of 3950\AA$\,$ in the [OII] spectrum and from
$\sim\,$4800$\,-\,$5100$\,$\AA~in the [OIII] spectrum was compared
with the flux blueward of 3800\AA$\,$.  Unlike Winge~et.$\,$al.,
(1998), we do not see an increased blue continuum for the near nuclear
clouds in the background subtracted image.  Given the complex
near-nuclear environment in our slitless spectrum and the weak
increase in the blue continuum seen in the background un-subtracted
spectrum it is possible that the background subtraction has masked a
weak increase in the blue continuum.

\section{Discussion}
\label{sec-discussion}

As noted in Section~\ref{sec-results}, our measurements
(Figure~\ref{rv}) suggest that there are at least two distinct
kinematic components. One component, the LVLVD clouds, is
characterized by low velocities
($\mid$v$\mid\,<\,$400$\,$km$\,$s$^{-1}$) and low velocity dispersions
($\Delta\,v\,<\,$130$\,$km$\,$s$^{-1}$).  A second cloud population,
the HVHVD clouds, appears to be characterized by both high velocities
and high velocity dispersions
(400$\,<\,\mid$v$\mid\,{^{<}_{\sim}}\,$1700$\,$km$\,$s$^{-1}$,
$\Delta\,v\,\geq\,$130$\,$km$\,$s$^{-1}$).  The HVHVD cloud population
is spatially confined to the inner 1.1$\arcsec$ ($\sim\,$70$\,$pc) of
the NLR, whereas the LVLVD clouds are distributed from 0.2$\arcsec$ to
$\sim\,$19$\arcsec$.

In this section we compare dynamical models with the observed
kinematic distribution by evaluating the projected radial velocity
distribution in the NLR in terms of rotation about the nucleus,
jet-cloud interactions, and wind driven outflows. Then we discuss the
relevance of the flux measurements to both dynamical and
photoionization models.

The radial velocity distribution of the NLR clouds (Figure~\ref{rv})
is reminiscent of the ``S-shape'' signature of gravitational rotation
about a massive central source (e.g.,~\cite{bower98}).  To investigate
the significance of this distribution, we have compared the measured
velocity distribution with the velocity profile expected from a simple
disk undergoing Keplerian rotation.

The inclination of the model disk was determined by the ratio of the
``minor'' to ``major'' axis in the inner NLR .  This yields an
inclination of 70$^{\circ}$ (Figure~\ref{o3a}).  In generating this
model, we have sampled the model at the same positions as the [OIII]
velocity measurements. Since the [OIII] peak is known to be
blueshifted by $\sim\,$40 km$\,$s$^{-1}$ relative to neutral hydrogen,
we have displaced the observed velocities by $\sim
-$40$\,$km$\,$s$^{-1}$.  Imposing this shift improves the agreement
between the model and the data (Figure~\ref{rv}). A central mass of
$\sim\,$4$\,\times\,$10$^8$M$_{\odot}$ is required by the model.

However, as illustrated in Figure~\ref{rv}, the data are not well fit
by the model.  Obviously there is a population of high velocity clouds
within a 1.1$\arcsec$ ($\sim\,$70$\,$pc) radius which do not fit the
model at all. Including these clouds in the model implies a
prohibitively high central mass of 10$^{10}\,$M$_{\odot}$. In
addition, the transition between redshifted and blueshifted velocities
near the nucleus is too soft to be indicative of gravitational
rotation under the influence of a central black hole.  The model
exhibits a sharp transition at the nucleus which is not mirrored by
the data.  Furthermore, the geometry implied by the model requires
that the axis of rotation of the disk is perpendicular to the radio
jet axis, a geometry that would be difficult to explain.  A recent
analysis (\cite {winge99}) of [OIII]$\,\lambda$5007 FOC spectra
augmented with both STIS slitless spectroscopy results (\cite
{hutchings98}) and ground based data interprets the NLR gas kinematics
within the central 4$\arcsec$ radius as being governed by Keplerian
rotation about an extended central mass of
$\sim\,$10$^{9}$M$_{\odot}$. However, our analysis indicates that the
high velocity kinematic component co-existing in this region is not
the result of simple gravitational motion.  

Other gravitationally dominated models involve orbits which may
manifest themselves as bars or triaxial structure.  The degree of
symmetry across the nucleus requires any of these to be rather
specially oriented with respect to our line of sight, in order to
produce the observed radial velocities.  In addition, all
gravitational models have to be oriented such that the observed cloud
morphology and velocity structure are not connected to the radio axis
which is generally considered to be aligned by the angular momentum
vector of the innermost material around the central mass.  We
therefore conclude that simple gravitational models cannot explain the
entire observed velocity distribution within $\sim\,$3$\arcsec$ of the
nucleus.

At distances exceeding 4$\arcsec$ ($\sim\,$260$\,$pc) from the
nucleus, we find that the gas cloud kinematics are consistent with
circular rotation in a gravitational field.  This is in agreement with
previous {\it {HST}} and ground based results (e.g.~\cite{penston90},
\cite {robinson94}, \cite {vila-vilaro95}, \cite {hutchings98}, \cite
{winge99}).

As noted in the previous section, we find no detailed correspondence
between the optical clouds and the radio structure.  However, the
general distribution of HVD clouds about the radio axis may indicate
that the radio jets are connected with the driving mechanism for these
clouds and may be influential in producing their high velocity
dispersions.  This conclusion is consistent with previous results
(\cite {winge99}) which find that the cloud kinematics within
4$\arcsec$ of the nucleus are disturbed but not dominated by
interactions with the radio jet.

The [OIII] flux does not appear to be enhanced for clouds near the
radio axis, nor does [OIII]/H$\beta$ appear to be a function of radio
axis location.  From our data we see no evidence that the jet has
affected the ionization state of the NLR gas.

Since neither gravitational rotation nor jet-cloud interactions are
successful in describing the observed kinematic distribution of the
clouds within $\sim\,$3$\arcsec$ of the nucleus, another physical
mechanism, such as radiation or wind driven outflow, must be
responsible for the gas kinematics.  In Hutchings et~al., (1998), we
argued against infall based upon opacity and similarity arguments.
These arguments still hold.


In ground based studies with lower spatial resolution, blue
asymmetries have been observed in the line profiles (e.g.~\cite
{heckman81}, \cite{whittle85}, \cite{dahari88}).  Dust extinction
within the NLR has been invoked to explain this asymmetry.  From our
observations, the blueshifted clouds SW of the nucleus appear brighter
and more extended than the redshifted clouds NE of the
nucleus. Assuming the SW cone is on the far side of the nucleus, our
observations are consistent with infalling, intrinsically dusty
clouds.  If the NE cone is on the far side of the nucleus, then our
observations are consistent with outflowing clouds in a dusty
intercloud medium.  The outflow scenario is compelling for two
reasons.  First, all the absorbers in the absorption line spectra of
C$\,$IV have outflow velocities (\cite{weyman97},
\cite{hutchings98}). Second, for the high velocities in the inner NLR,
gravitationally driven infall requires an improbably high central mass
of $\sim\,$10$^{10}\,$M$_{\odot}$. Furthermore, radio observations
(\cite{pedlar93}, \cite{mundell95}, \cite{ulvestad98}) map an
outflowing radio jet that is roughly aligned with the optical axis of
the NLR bicone.


Emission line ratios of the individual clouds are consistent with
photoionization models.  The [OIII]/H$\beta$ ratio declines roughly
linearly within an inner $\sim\,$3$\arcsec$ (200$\,$pc) radius from
the nucleus, particularly among the HVD clouds.  This decrease in the
flux ratio indicates a decreasing ionization parameter
(\cite{ferland83}). Since the ionization parameter is proportional to
r$^{-2}$n$^{-1}$, it appears that the density, n, must decrease as
$\sim\,$r$^{-1}$.  At distances further from the nucleus where the
clouds have low velocity dispersions, the [OIII]/H$\beta$ ratio is
roughly constant.  Within the scatter the [OIII]/H$\beta$ ratio in
this region is comparable to its initial value, indicating that either
the ionization parameter is roughly constant at larger radii with the
density decreasing as r$^{-2}$ or the data are subject to projection
effects, local gradients, or both.

The ratio of [OI]/[SII] should also be a density indicator.  However,
the small sample of clouds for which we have flux measurements and the
uncertainties in these measurements could mask a change in the flux
ratio as a function of radial distance.  The ratios listed in Table~5
are consistent with a constant of 0.35 for the 120$\,$pc radius
spanned by these measurements.

\section{Conclusions}

We have spatially mapped the full velocity field of the NLR for
NGC$\,$4151.  Our data support a biconical distribution of emission
line gas where clouds to the SW of the nucleus are approaching and
clouds to the NE have recessional velocities.  Within an inner
3.2$\arcsec$ radius we observe a high velocity dispersion
($\Delta$v$\,\geq\,$130$\,$km$\,$s$^{-1}$) population of clouds. The
[OIII]/H$\beta$ flux ratio declines roughly linearly for these clouds.
Closer to the nucleus, and distributed within a cone of opening angle
$\sim\,$45$^{\circ}$ about the NLR axis, is a population of faint
HVHVD clouds.

Some dynamical scenarios have been considered to account for the
observationally derived kinematic distribution, velocity profiles, and
the radial dependence of the flux ratios arising in the NLR.
Kinematic mapping of the NGC$\,$4151 NLR indicates that more than one
process must be employed to model the entire velocity field associated
with the NLR emission.  Our data are consistent with photoionization
by a central source and outflow from a spatially unresolved nuclear
region.  Within an inner 3.2$\arcsec$ radius we do not find evidence
to support gravitationally dominated models such as rotation about a
central massive source or transport along a nuclear bar structure.

The evidence we find for radio jet-cloud interactions is provocative
but weak. While the HVD clouds are relatively symmetrically
distributed within a 45$^{\circ}$ opening angle cone centered on the
NLR axis at $\sim\,$63$^{\circ}$, the LVLVD and HVHVD clouds are
co-spatial within the cones and there is no preference for the HVD or
HVHVD clouds to be associated with the radio knots which lie along a
PA of 77$^{\circ}$.

For the HVD clouds the velocity distribution is sharply peaked about
the nucleus and the [OIII]/H$\beta$ flux ratio declines approximately
linearly.

Our results are consistent with a wind driven outflow as the
acceleration mechanism for these clouds. Qualitatively the HVHVD
clouds resemble the tenuous cloud component that is being radially
driven outward in NGC$\,$1068 (\cite {cecil90}).  The LVHVD cloud
population resident within the same projected radius may be
participating in the same wind driven outflow but its projected
velocity along our line of sight is small.  Alternate driving
mechanisms for these clouds are less plausible. Radiative acceleration
is inefficient, and gravitationally driven infall requires an
improbably high central mass of
$\sim\,$4$\,\times\,$10$^{9}\,$M$_{\odot}$ for radial velocities of
800$\,$km$\,$s$^{-1}$.

Recent {\it {HST}} STIS observations by this collaboration have mapped
the NGC$\,$4151 emission line spectra from 1150$\,$\AA~through
1.0$\mu$m at two slit positions which intersect multiple clouds in the
NLR.  We will use these data as input to photoionization models to
probe the density and ionization structure within the NLR.

We thank Carole Mundell for providing us with an electronic version of
the radio data (\cite {mundell95}) which is presented in
Figure~\ref{radio_map}.  Support for this project was provided in
response to NASA Announcement of Opportunity OSSA-4-84 through the
{\it {Hubble Space Telescope}} Project at Goddard Space Flight Center.

\vspace{0.2truein}


\clearpage

\normalsize

\normalsize

\figcaption{Dispersed G430M image from STIS of the blue region of the
spectrum of NGC$\,$4151. The positions of the main features in the
nuclear spectrum are labelled. The upper panel is background
subtracted, to enable the faint inner emission clouds to be seen. The
lower panels show the spectrum as observed, with two stretches to show
the inner and outer clouds. The wavelength scale is in the observed
frame. 
\label{blue}}

\figcaption{Dispersed G430M image from STIS in the observed frame of
the H$\beta$ and [O III] spectral region of NGC$\,$4151. This may be
compared with the image of the same region with different spectral
dispersion direction, shown in Hutchings et al., 1998.
\label{green}}

\figcaption{Dispersed G750M image from STIS of the red region of the
spectrum.  Two stretches are shown to cover the full dynamic range of
cloud fluxes.  The wavelength scale is in the observed
frame. \label{red}}
    
\figcaption{Direct WF image in the light of [O III] (F502N), with the
orientation of the later observations, to identify the outer clouds
measured. 
\label{o3b}}
           
\figcaption{Direct image of inner clouds, labelling them for
identification.  The image is in the light of [O III] taken with the
PC and filter F502N.  The image is rotated to the orientation of the
initial STIS dispersed images. The dashed ellipse illustrates
the inclination angle used in modelling the cloud velocity
structure. \label{o3a}}
 
\figcaption{Dispersed [O III] 5007$\,$\AA~line image from the STIS,
labelling the inner clouds measured. The dispersion direction is L-R,
and the nuclear and galaxy continuum has been subtracted to make
individual cloud features more visible. Cloud velocities are given in
Tables 2 and 3. \label{o3d}}
          
\figcaption {This diagram illustrates how cloud locations and
velocities are obtained with two different roll angles. The thick
lines represent the nuclear continuum, and show the two dispersion
directions that occur because of the different roll angles.  The
images are aligned at the nucleus with the known rotational angle
differences. In the two dispersed images, we see the continuum and
emission line cloud images positioned along the same dispersion
direction, determined by both their spatial location and their radial
velocity. The undispersed cloud position is located where the lines
meet when extended along the two dispersion directions. For example,
cloud A appears at A1 in one image and A2 in the other. The location
of A is where these two lines intersect, and this is shifted by the
same distance (in the geometrically corrected image) A-A1 and
A-A2. The four clouds illustrated show the different relative
positions they will have with the two roll angles shown.
\label{roll}}

\figcaption{Radial velocity of clouds with distance from the nucleus
perpendicular to the dispersion direction (vertical in
Figure~\ref{o3b}: approximately along the radio axis). In the upper
panel, the two symbols distinguish the broad and narrow lines. The
plot is not substantially changed if the projected distance from the
nucleus is used. The lower panel shows the velocity curve for the
inclined disk model, with the radial velocities shifted by -40
km$\,$s$^{-1}$, as discussed in the text.  In the lower panel a model
for a rotating thin disk about a central mass of
4$\,\times\,$10$^{8}$M$_{\odot}$ is denoted by a dashed
line. \label{rv}}

\figcaption{Cloud velocity dispersion with distance from the nucleus
perpendicular to the dispersion direction (vertical distance in
Figure~\ref{o3b}). The high velocity clouds are those which have
radial velocities in excess of 400 km$\,$s$^{-1}$. \label{width}}

\figcaption{Histogram of cloud velocity dispersions, showing how the
bin values change for inner subsets of clouds. In all cases the
distribution appears bimodal, which we discuss as two populations of
clouds with dispersions above and below 130
km$\,$s$^{-1}$. \label{hist}}

\figcaption{The upper panel illustrates the spatial distribution of
the [OIII] emission line clouds within 3.2$\arcsec$ of the
nucleus. The population of HVD, HVHVD, and LVD clouds are denoted,
Position angles of the radio cores relative to the nucleus (\cite
{pedlar93}, \cite {mundell95}), the stellar bar (\cite
{pedlar92}), and the extended NLR are illustrated. The radio knots C5
and C2, which define a position angle of
73$^{\circ}$, are depicted as open squares.\label{radio}}

\figcaption{The MERLIN radio map (\cite {mundell95}) overlaid on the
F502N PC image.  While the NLR [OIII] emission is globally aligned
with the radio emission, there is a lack of detailed correspondence
between the [OIII] clouds and the radio knots.  The resolution of the
radio map is 0.15$\,\times\,$0.16 arcsec. The radio contour levels are
at [1,2,3,4,5,6,7,8,9,10,15,20,25,30]$\,\times\,$1.7$\,$mJy and
0.96$\,$mJy, where the 5$\,\sigma$ contour is at 1.7$\,$mJy and the
3$\,\sigma$ contour is at 0.96$\,$mJy. \label{radio_map}}

\figcaption{Line flux measures and ratios for clouds.  The symbols
distinguish approaching and receding clouds in the upper two and lower
left panel. In the lower right panel the symbols distinguish between
high and low velocity dispersion clouds.  Note the decline in
[OIII]/H$\beta$ in the inner region, which is seen primarily in the
lines with high velocity dispersion (lower right panel).
\label{fluxes}}

\figcaption{Line ratio dispersed image showing [OIII]/H$\beta$. Galaxy
and scattered nuclear continuum light have been removed before
deriving the ratio. Dark indicates high ratio, as seen in the outer
and the high velocity clouds. \label{ratio}}

\clearpage

\begin{deluxetable}{lcllrrr}
\tablenum{1}
\tablewidth{6in}
\tablecaption{NGC$\,$4151 data log}
\tablehead{\colhead{Date} & \colhead{Instrument} &\colhead{Filter}
&\colhead{Grating} &\colhead{$\lambda_c$} &\colhead{$\Delta\lambda$} &\colhead{Exposure}
\\ 
& & & &[\AA] & [\AA] & Time [s]}
\startdata 
1995 January 23\tablenotemark{a}& WFPC2&   F502N&    & 5012&  26.8&  860 \nl
1997 March 29\tablenotemark{a}& STIS & Clear & G430M & 4961& 282  & 2690 \nl
1997 July 15& STIS & Clear & G430M & 3843& 284 & 3390 \nl
1997 July 15& STIS & Clear & G430M & 4961& 282 & 2139 \nl
1997 July 15& STIS & Clear & G750M & 6581& 569 & 1960 \nl
1997 July 15& STIS & [OII] &       & 3740&  80 & 3070 \nl
1997 July 15& WFPC2& F502N &       & 5012&  26.8& 1800 \nl
1998 April 5& WFPC2& F390N &       & 3889&  45.3& 1200 \nl
1998 April 5& WFPC2& F656N &       & 6562&  22 &  700 \nl
1998 April 5& WFPC2& F658N &       & 6590&  28.5& 402 \nl
1998 April 5& WFPC2& F673N &       & 6733&  47.2& 1200 \nl
\enddata
\tablenotetext{a}{Archival data}
\end{deluxetable}

\clearpage

\begin{deluxetable}{lrrrr}
\tablenum{2}
\tablewidth{5in}
\tablecaption{\label{tbl-velocities} Cloud Velocities}
\tablehead{\colhead{Cloud \#} &\colhead{Radial Velocity\tablenotemark{a}} &\colhead{$\sigma_{RV}$}
&\colhead{$\Delta\,v$ (Dispersion)} &\colhead{$\sigma_{\Delta\,v}$}\\
&km$\,$s$^{-1}$  & km$\,$s$^{-1}$   & km$\,$s$^{-1}$  &km$\,$s$^{-1}$}
\startdata 
1         &230  &24 &183 &74 \nl
2         &47   &25 &340 &14 \nl
3         &66   &40 &63  &13 \nl
4         &278  &14 &44  &27 \nl
5         &192  &12 &75  &5 \nl
6         &11   &13 &153 &38 \nl
7         &105  &-  &94  &- \nl
8         &265  &18 &181 &24 \nl
9         &72   &10 &83  &20 \nl
9b        &10   &16 &241 &4  \nl
10        &495  &21 &160 &45 \nl
11        &36   &13 &223 &7 \nl
12        &337  &36 &300 &14 \nl
13        &190  &14 &24  &21 \nl
14        &125  &21 &205 &6 \nl
15        &-93  &1  &282 &36 \nl
16        &-152 &77 &317 &115 \nl
17        &-107 &17 &248 &128 \nl
18        &-235 &47 &157 &38  \nl
19        &-122 &29 &396 &43 \nl
19a       &-215 &-  &79  &- \nl
20        &36   &19 &78  &12 \nl
21\tablenotemark{b}       &-538  & 10&294 &- \nl
21x\tablenotemark{c}       &-43  &-  &88  &- \nl
22        &-279 &19 &116 &6 \nl
23        &-280 &24 &198 &45 \nl
24        &NC\tablenotemark{d}   & & & \nl
25        &-248 &16 &243 &10 \nl
26        &-814 &53 &304 &4 \nl
27        &-161 &29 &119 &169 \nl
28        &190  &32 &156 &141 \nl
29        &84   &21 &70  &71 \nl
30        &78   &15 &13  &14 \nl
31        &20   &-  &210 &- \nl
32        &ghost&   &    & \nl
33        &7    &21 &64  &55 \nl
34        &7    &26 &43  &16 \nl
35        &7    &14 &48  &-  \nl
36        &-3   &8  &76  &25 \nl
37        &24   &13 &34  &9 \nl
38        & 27  &14 &46  &10 \nl
39        &6    &25 &41  &37 \nl
40        &0    &1  &51  &-  \nl
41        &8    &18 &59  &51 \nl
42        &-62  &9  &68  &63 \nl
43        &-7   &-  &74  &- \nl
44        &5    &-  &0   &- \nl
45        &-18  &-  &0   &- \nl
46        &-30  &-  &0   &- \nl
47        &-5   &-  &45  &- \nl
48        &4    &-  &0   &- \nl
49        &-13  &-  &101 &- \nl
50        &-20  &-  &0   &- \nl
51        &-20  &-  &0   &- \nl
\enddata
\tablenotetext{a}{Relative to the heliocentric recessional velocity of
997$\,$km$\,$s$^{-1}$ (\cite {pedlar93})}
\tablenotetext{b}{Velocity is from earlier epoch (\cite{hutchings98}), 
this cloud is blended with nucleus in this epoch.}
\tablenotetext{c}{Closest UNblended cloud to the nucleus}
\tablenotetext{d}{Not Confirmed}
\end{deluxetable}
\clearpage

\begin{deluxetable}{lrccrrc}
\tablenum{3}
\tablewidth{5in}
\tablecaption{High Velocity Clouds}
\label{tbl-highv}
\tablehead{\colhead{Cloud \#} &\colhead{RV} &\colhead{Uncertainty$_{RV}$}
&\colhead{$\sigma_{\Delta\,v}$} &\colhead{dx$_{\circ}$} &\colhead{dy$_{\circ}$} & Neighboring Cloud\\
&km$\,$s$^{-1}$ &km$\,$s$^{-1}$  & km$\,$s$^{-1}$   & {\arcsec}  & {\arcsec} & }
\startdata 
a            &-1716 & 256  & 175 &0.493 &-0.963 & 19     \nl
b            &-1516 & 261  & 140 &0.028 &-0.963 & 22     \nl
c            &-1107 & 187  & 278 &0.231 &-0.913 & 19, 19a\nl
d            &-1448 & 248  & 291 &0.606 &-0.659 & 15     \nl
n            &-1162 & 412  & 393 &0.078 &-0.254 & 21x, 20 \nl
g            &846   & 104  & 265 &-0.253 &0.811 &  8     \nl
\enddata
\end{deluxetable}
\clearpage

\begin{deluxetable}{cccc}
\tablenum{4}
\tablewidth{5in}
\tablecaption{H$\beta$ and [OIII] Cloud Fluxes Derived From Gaussian Fits}
\tablehead{\colhead{Cloud \#} &\colhead{H$\beta$ Flux}& \colhead{[OIII] Flux} &\colhead{[O III]/H$\beta$}\\
 & erg s$^{-1}$ cm$^{-2}$ & erg s$^{-1}$ cm$^{-2}$ & }
\startdata 
1       & 3.09$\times10^{-14}$  & 3.29$\times10^{-13}$  & 10.65 \nl
2       & 1.35$\times10^{-14}$  & 1.42$\times10^{-13}$  & 10.52 \nl
3       & 8.91$\times10^{-16}$  & 9.71$\times10^{-15}$  & 10.90 \nl
4       & 4.92$\times10^{-15}$  & 4.90$\times10^{-14}$  &  9.96 \nl
5       & 2.76$\times10^{-15}$  & 3.14$\times10^{-14}$  & 11.38 \nl
6$+$11  & 2.92$\times10^{-14}$  & 3.37$\times10^{-13}$  & 11.54 \nl
7       & 3.45$\times10^{-15}$  & 3.45$\times10^{-14}$  & 10.00 \nl
8       & 8.33$\times10^{-15}$  & 9.77$\times10^{-14}$  & 11.73 \nl
9a      & 1.93$\times10^{-14}$  & 2.13$\times10^{-13}$  & 11.04 \nl
9b      & 2.30$\times10^{-14}$  & 2.45$\times10^{-13}$  & 10.65 \nl
10      & 1.77$\times10^{-14}$  & 1.57$\times10^{-13}$  &  8.87 \nl
12      & 1.30$\times10^{-14}$  & 1.48$\times10^{-13}$  & 11.38 \nl
13      & 2.38$\times10^{-15}$  & 2.12$\times10^{-14}$  &  8.93 \nl
14      & 3.02$\times10^{-15}$  & 2.17$\times10^{-14}$  &  7.19 \nl
15$+$17 & 2.77$\times10^{-14}$  & 3.41$\times10^{-13}$  & 12.31 \nl
16      & 1.09$\times10^{-14}$  & 1.25$\times10^{-13}$  & 11.47 \nl
18      & 1.06$\times10^{-14}$  & 1.02$\times10^{-13}$  &  9.62 \nl
19      & 4.58$\times10^{-14}$  & 4.60$\times10^{-13}$  & 10.04 \nl
19a     & 1.43$\times10^{-14}$  & 1.42$\times10^{-13}$  &  9.93 \nl
20      & 4.67$\times10^{-14}$  & 5.84$\times10^{-13}$  & 12.51 \nl
21x     & 2.64$\times10^{-14}$  & 2.91$\times10^{-13}$  & 11.02 \nl
22      & 2.28$\times10^{-14}$  & 3.07$\times10^{-13}$  & 13.46 \nl
23      & 3.27$\times10^{-14}$  & 3.53$\times10^{-13}$  & 10.80 \nl
25      & 1.72$\times10^{-14}$  & 1.46$\times10^{-13}$  &  8.49 \nl
26      & 7.11$\times10^{-15}$  & 1.21$\times10^{-13}$  & 17.02 \nl
27      & 6.17$\times10^{-15}$  & 5.78$\times10^{-14}$  &  9.37 \nl
28      & 2.90$\times10^{-15}$  & 2.03$\times10^{-14}$  &  7.00 \nl
29      & 7.89$\times10^{-16}$  & 1.19$\times10^{-14}$  & 15.08 \nl
30      & 9.43$\times10^{-16}$  & 1.31$\times10^{-14}$  & 13.89 \nl
35      & 3.98$\times10^{-16}$  & 5.70$\times10^{-15}$  & 14.32 \nl
37      & 5.90$\times10^{-16}$  & 8.26$\times10^{-15}$  & 14.00 \nl
39      & 3.89$\times10^{-16}$  & 4.54$\times10^{-15}$  & 11.67 \nl
41      & 2.93$\times10^{-16}$  & 3.12$\times10^{-15}$  & 10.65 \nl
42      & 5.20$\times10^{-16}$  & 6.86$\times10^{-15}$  & 13.19 \nl
\enddata
\end{deluxetable}

\tiny

\begin{deluxetable}{c||cc|cc|cc|cc|cc|cccc}
\tablenum{5}
\tablewidth{9.0in}
\tablecaption{NGC$\,$4151 Cloud Fluxes Derived From Gaussian Fits ($\times 10^{-14}$ erg s$^{-1}$ cm$^{-2}$)}
\tablehead{
\multicolumn{1}{c||}{Cloud} 
&\multicolumn{2}{c|}{[OII]}
&\multicolumn{2}{c|}{H$\beta$}
&\multicolumn{2}{c|}{[OIII]}
&\multicolumn{2}{c|}{[OI]}
&\multicolumn{2}{c|}{[SII]} 
&\colhead{[OII]/H$\beta$}
&\colhead{[OIII]/H$\beta$}
&\colhead{[OI]/H$\beta$}
&\colhead{[SII]/H$\beta$} \\
 &\multicolumn{1}{c}{Flux} 
 &\multicolumn{1}{c|}{Error}
 &\multicolumn{1}{c}{Flux} 
 &\multicolumn{1}{c|}{Error}
 &\multicolumn{1}{c}{Flux} 
 &\multicolumn{1}{c|}{Error}
 &\multicolumn{1}{c}{Flux} 
 &\multicolumn{1}{c|}{Error}
 &\multicolumn{1}{c}{Flux} 
 &\multicolumn{1}{c|}{Error}
 & & & &
}

\startdata 
1     &  4.11 & 0.49 & 2.60 & 0.69 & 26.9 & 6.32 & 1.36 & 0.58 & 2.51 & 0.21    & 1.58 & 10.35 & 0.52 & 0.97 \nl
6$+$11 &  2.93 & 0.91 & 1.55 & 0.33 & 19.3 & 4.22 & 1.18 & 0.16 & 3.08 & 0.51    & 1.89 & 12.45 & 0.46 & 0.82 \nl
9a    &  5.83 & 1.78 & 1.93 & 0.54 & 21.3 & 6.17 & 1.23 & 0.33 & 3.41 & 0.50    & 3.02 & 11.04 & 0.64 & 1.77 \nl
15    &  5.77 & 1.33 & 1.18 & 0.42 & 13.2 & 4.42 & 2.89 & 0.52 & 5.35 & $-$0.06    & 4.89 & 11.19 & 2.45 & 4.53 \nl
19    &  6.43 & 2.28 & 3.77 & 0.85 & 41.20 & 6.36 & 2.89 & 0.99 & 5.00 & 2.19   & 1.71 & 10.93 & 0.77 & 1.33 \nl
20    & 10.00 & 2.90 & 4.09 & 1.26 & 51.70 & 5.17 & 3.33 & 1.06 & 9.96 & 2.24   & 2.44 & 12.64 & 0.81 & 2.44 \nl
22    &  8.59 & 1.65 & 1.88 & 0.63 & 23.50 & 9.40 & 0.83 & 0.09 & 4.74 & $-$0.41   & 4.57 & 12.50 & 0.44 & 2.52 \nl
23    &  6.92 & 1.99 & 3.27 & 1.04 & 35.30 & 8.95  & 2.10 & 0.47 & 5.82 & 0.97   & 2.12 & 10.80 & 0.64 & 1.78 \nl
25    &  4.14 & 1.24 & 1.24 & 0.45 & 10.80 & 3.96  & 0.61 & 0.02 & 2.73 & 0.76 
& 3.34 & 8.71  & 0.49 & 2.20 \nl
30    &  0.55 & 0.12 & 0.12 & 0.04 &  1.65 & 0.53 &      &      &      &      & 4.62 & 13.87 &      &      \nl
\enddata
\end{deluxetable}

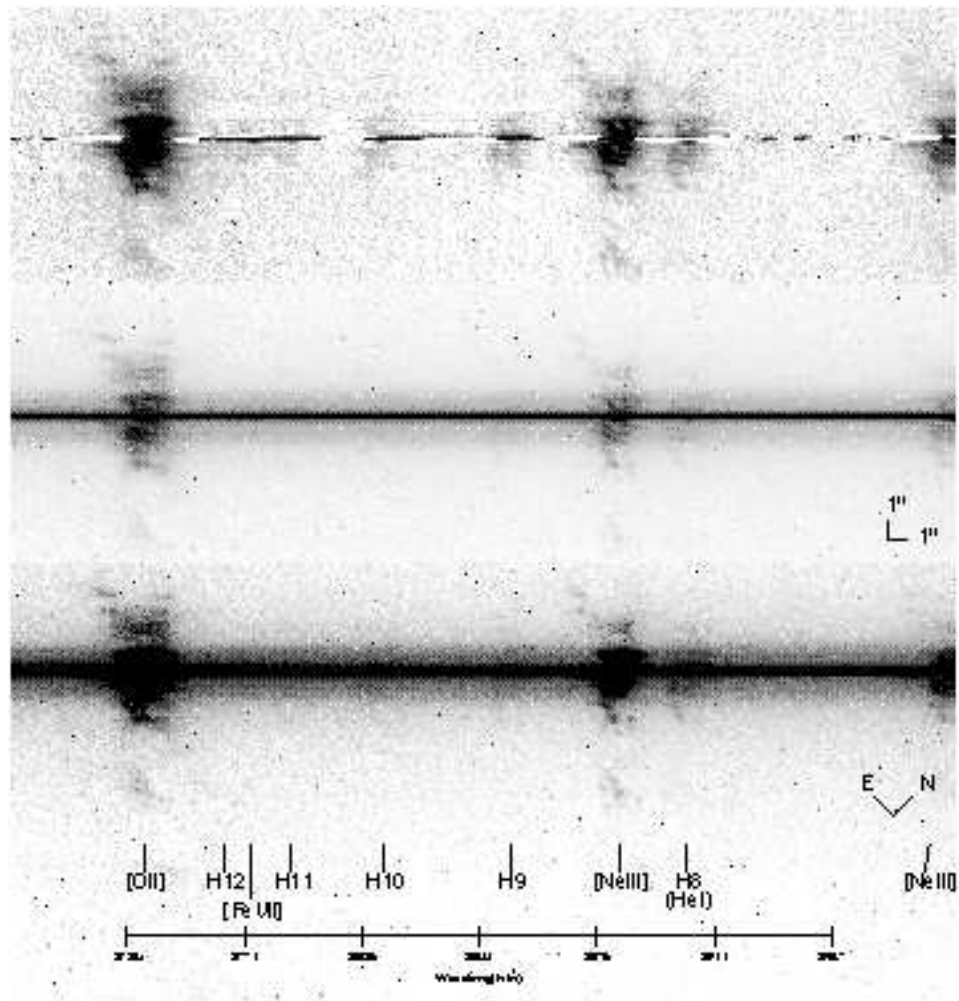
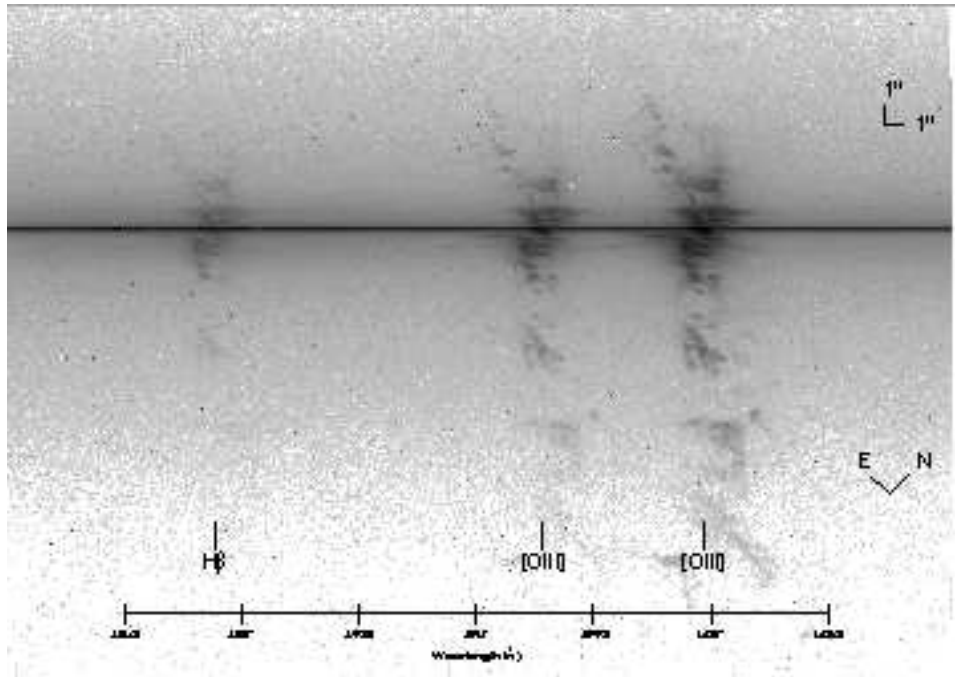
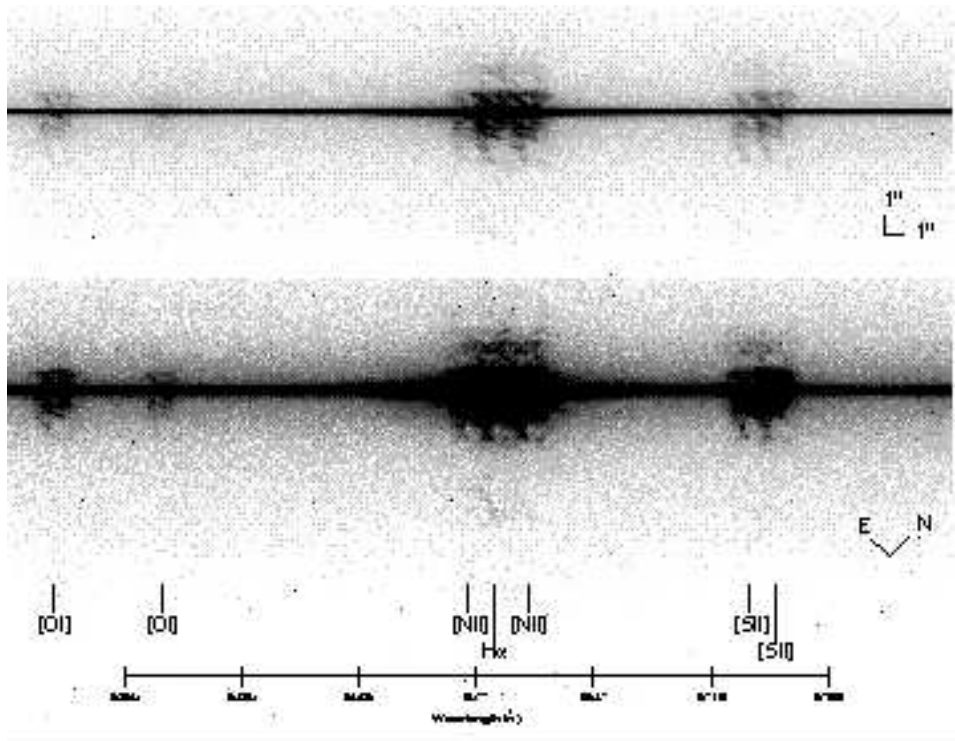
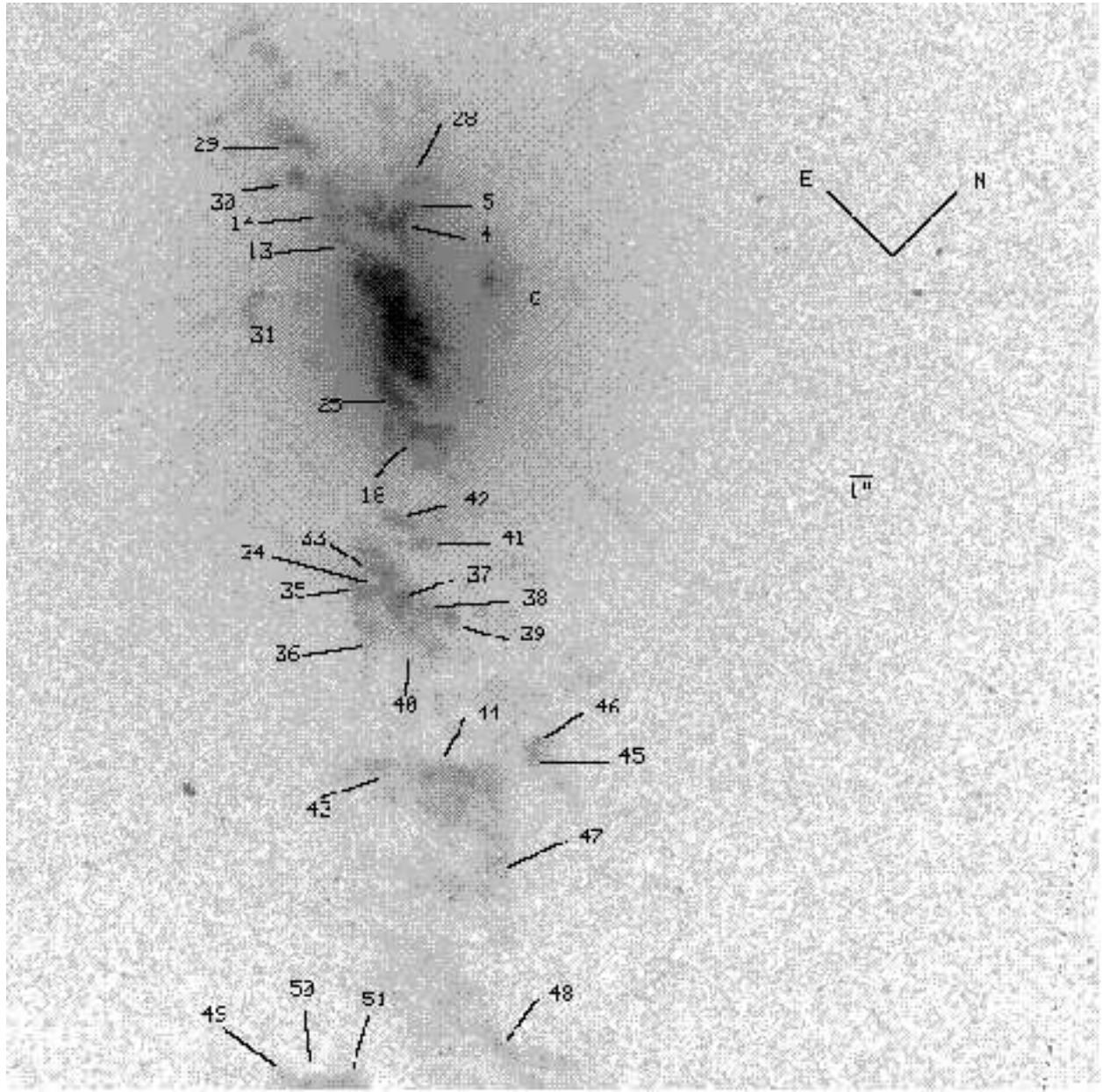
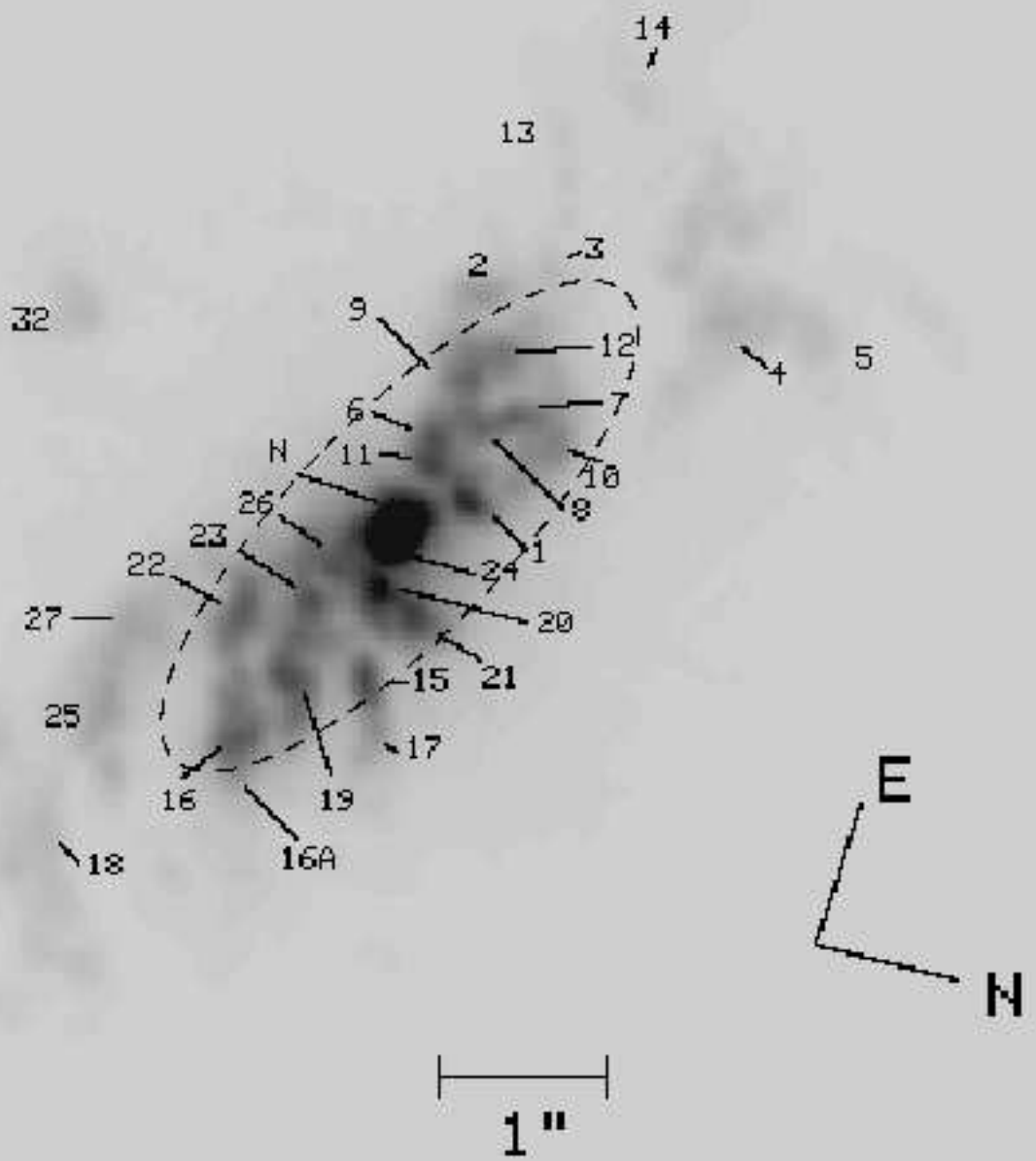
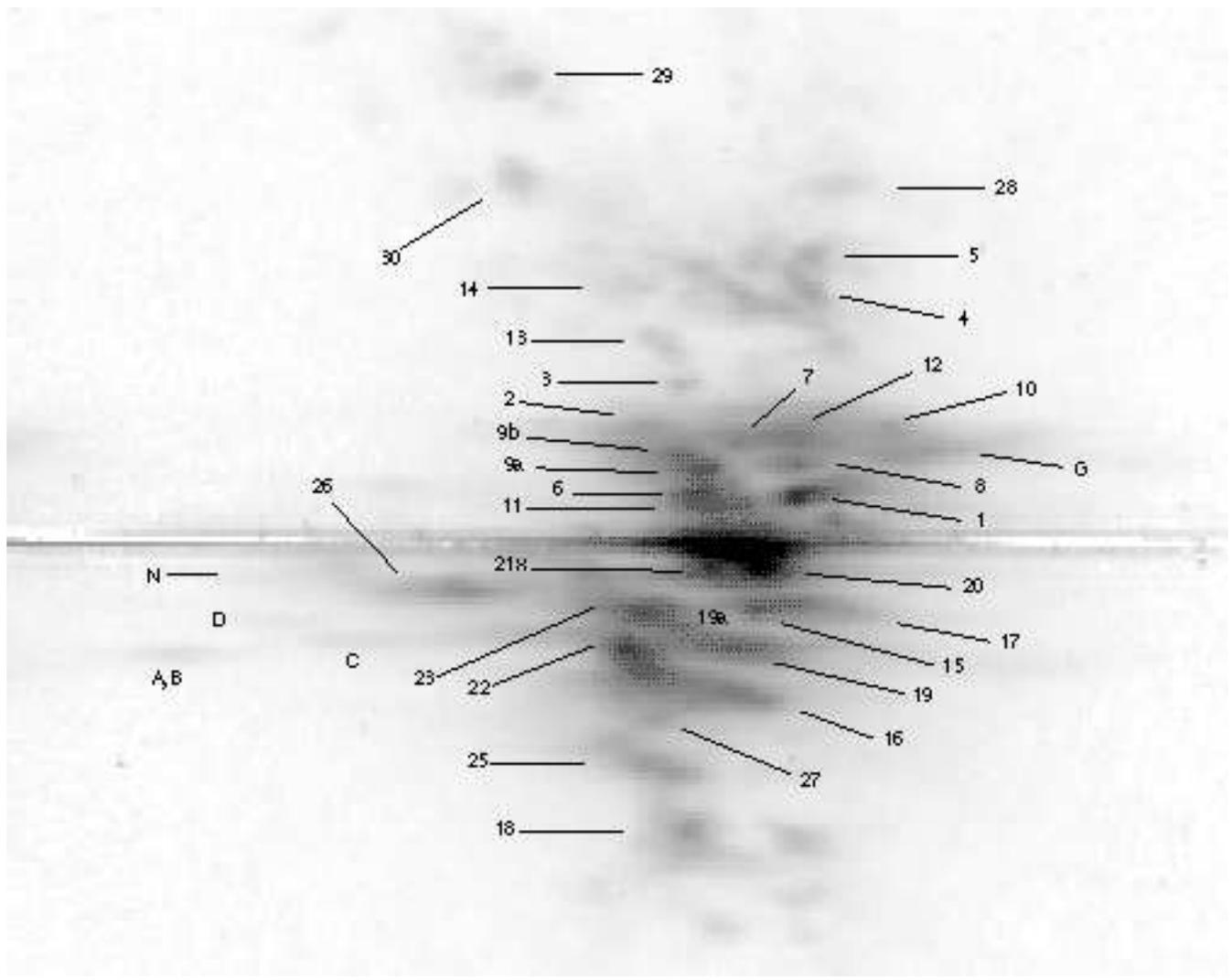
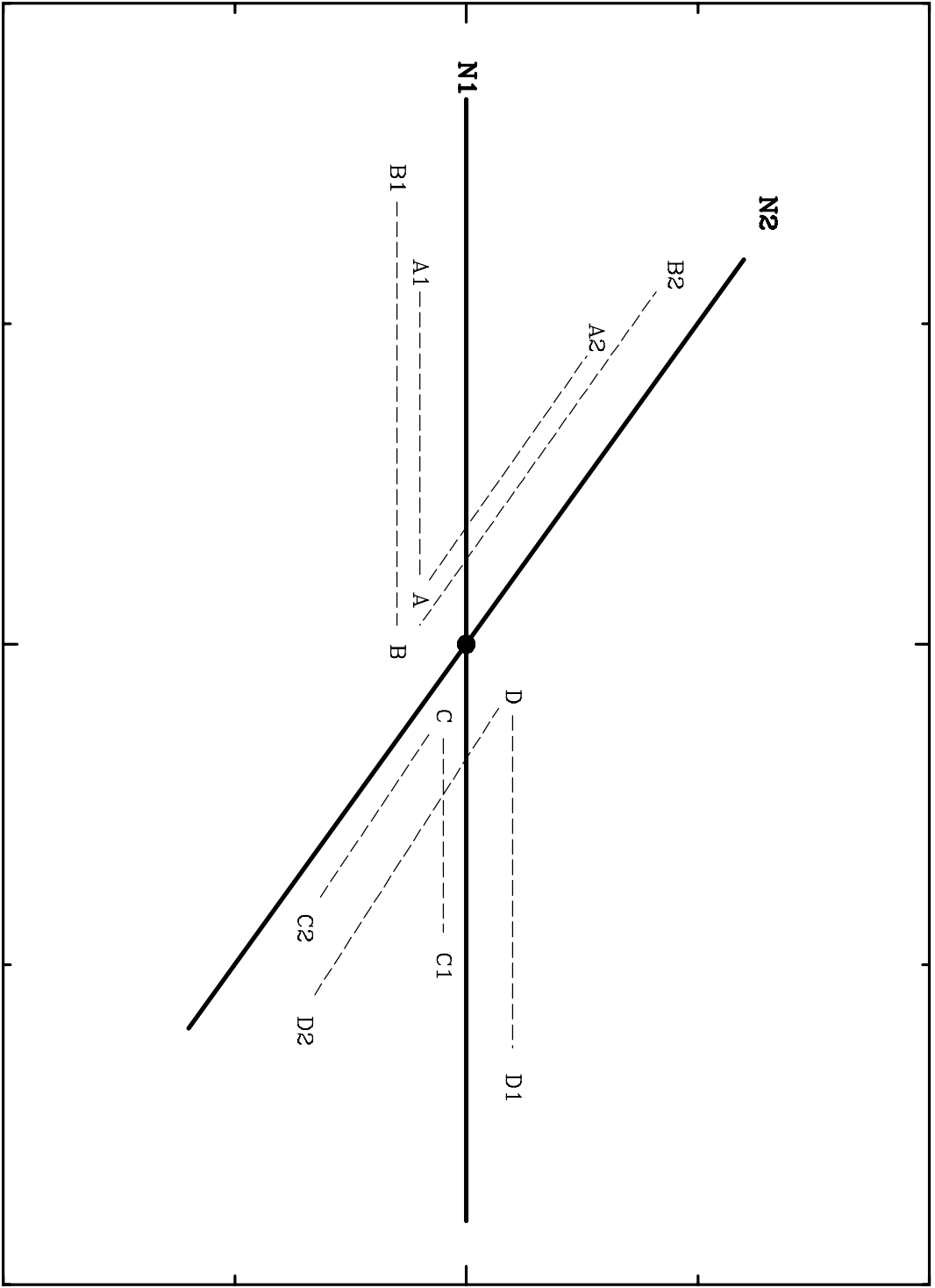
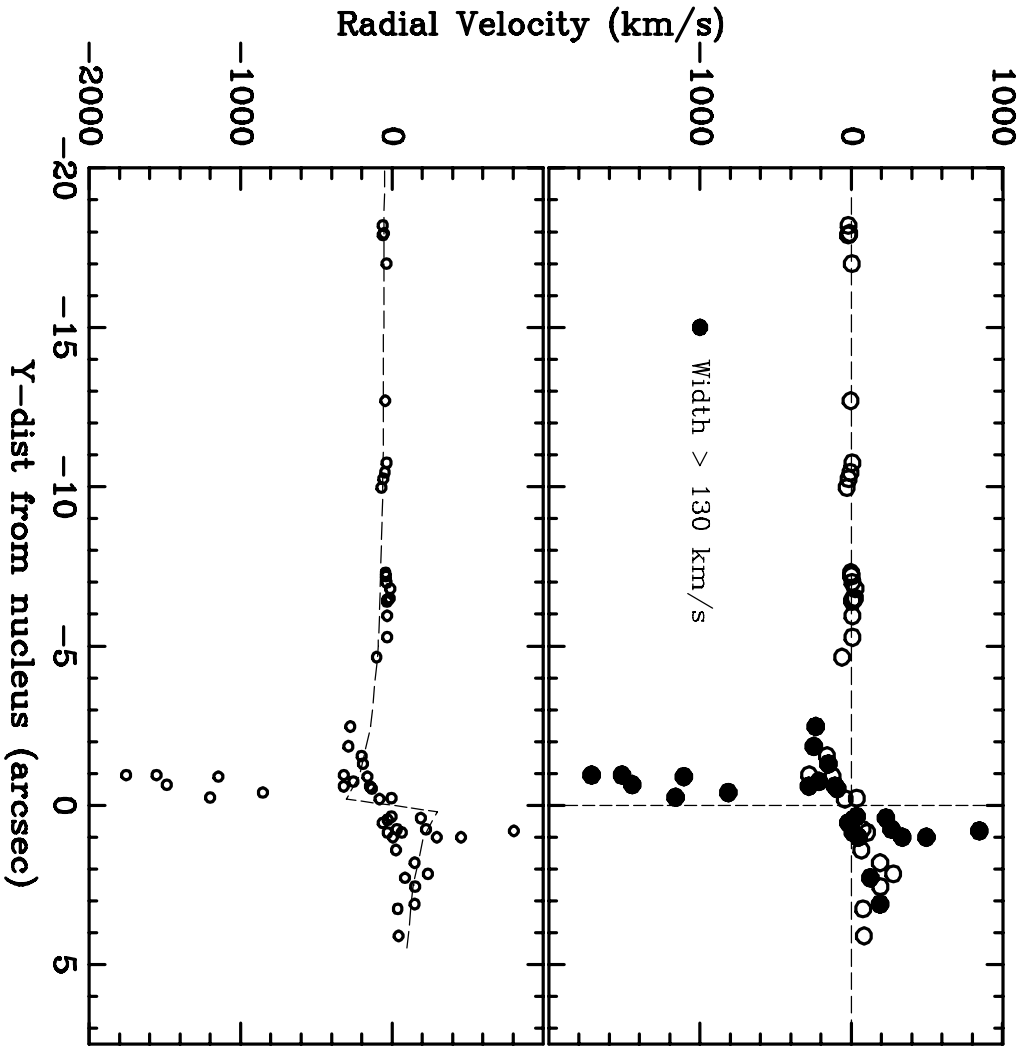
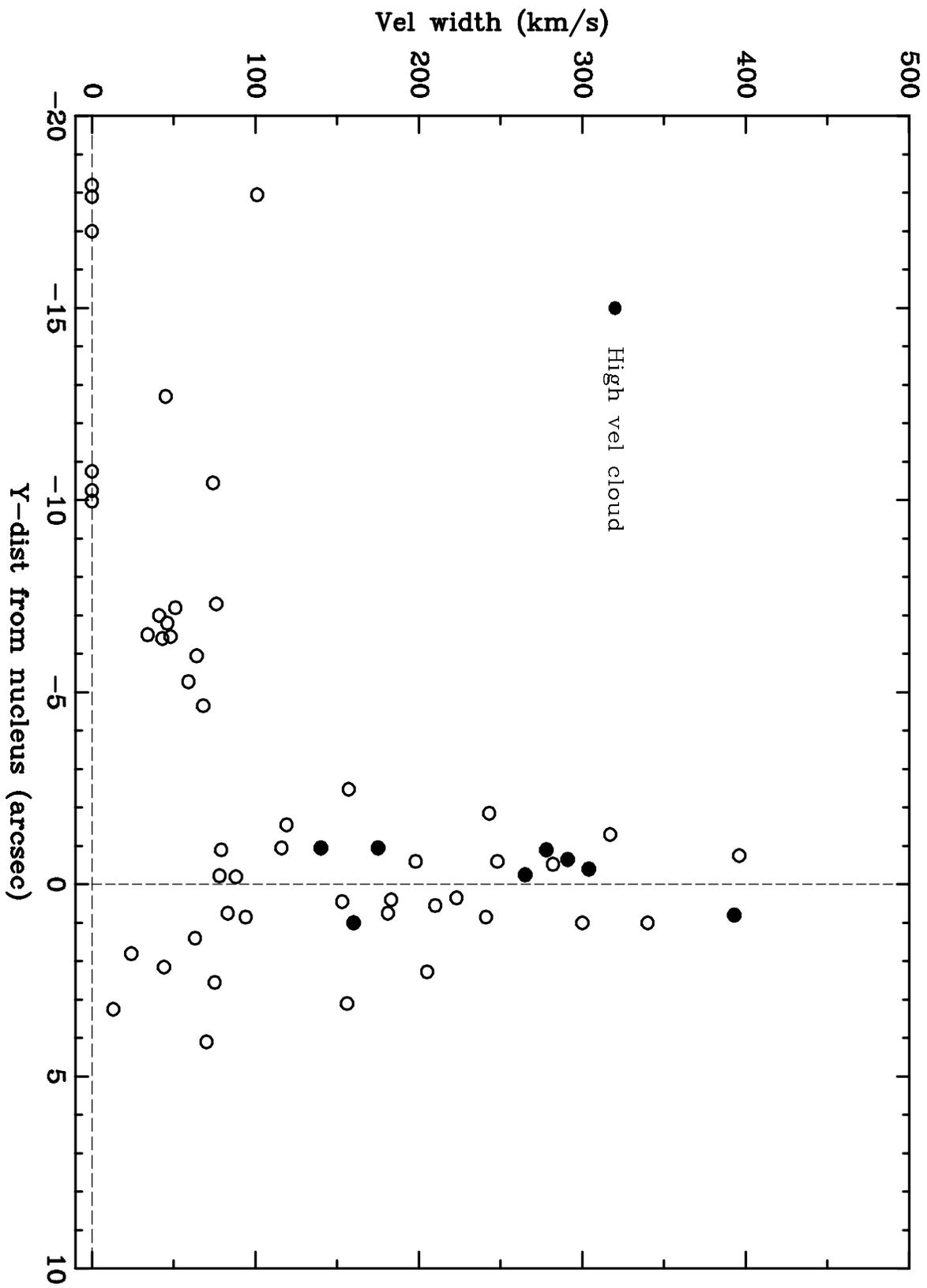
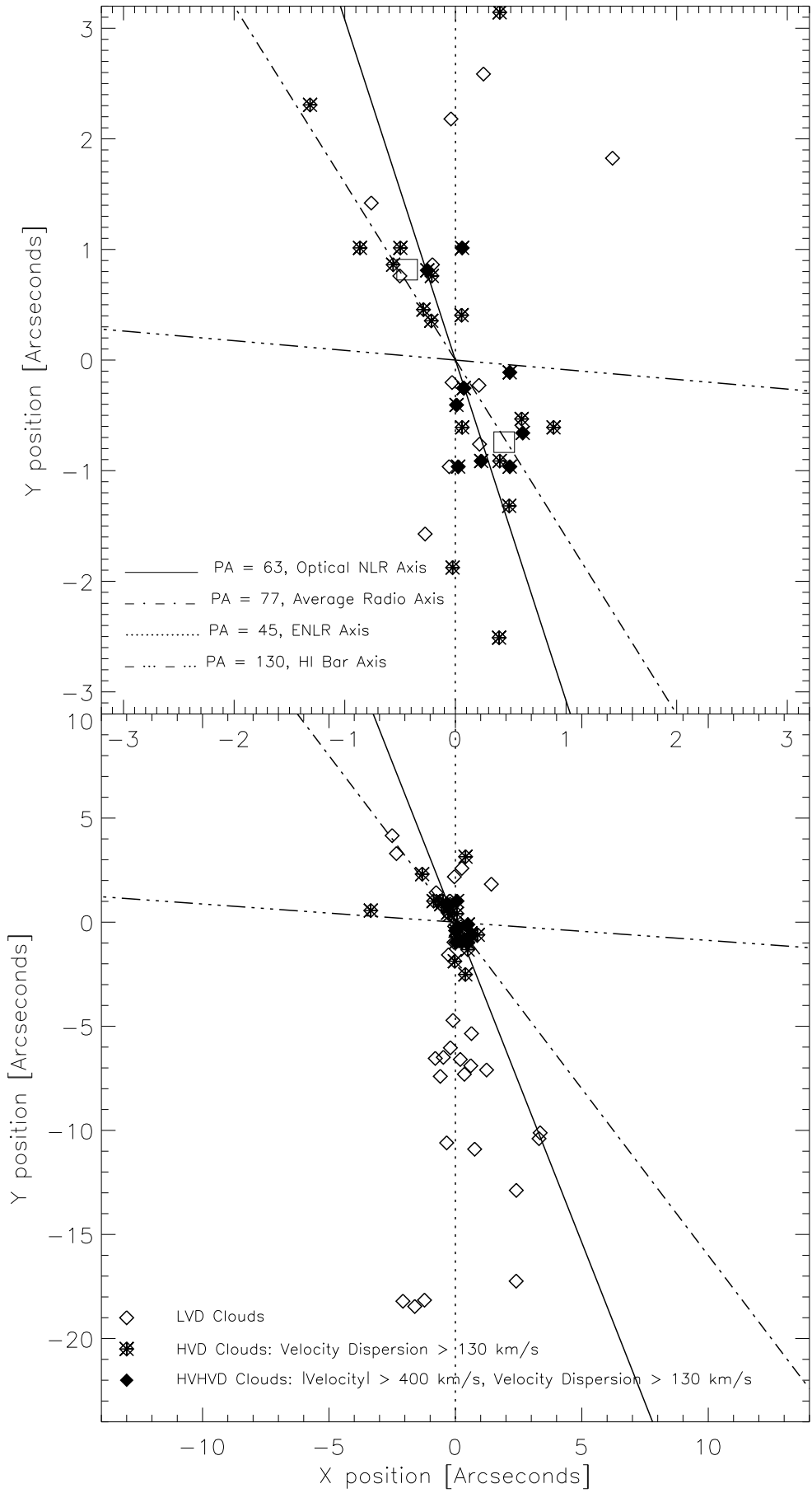
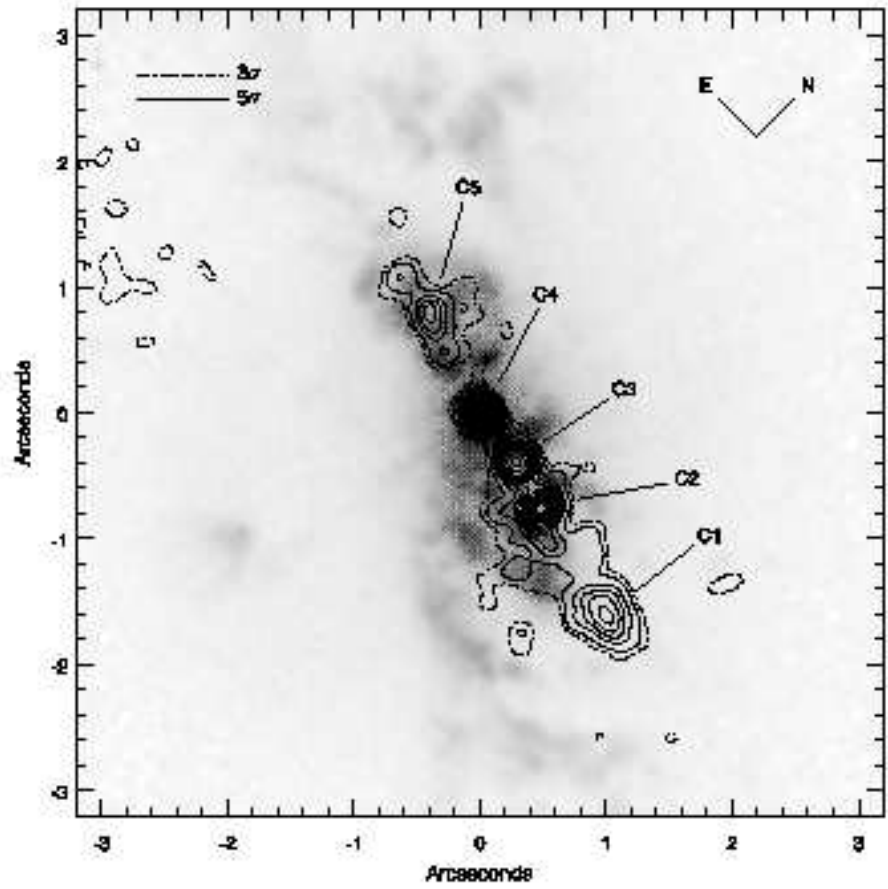
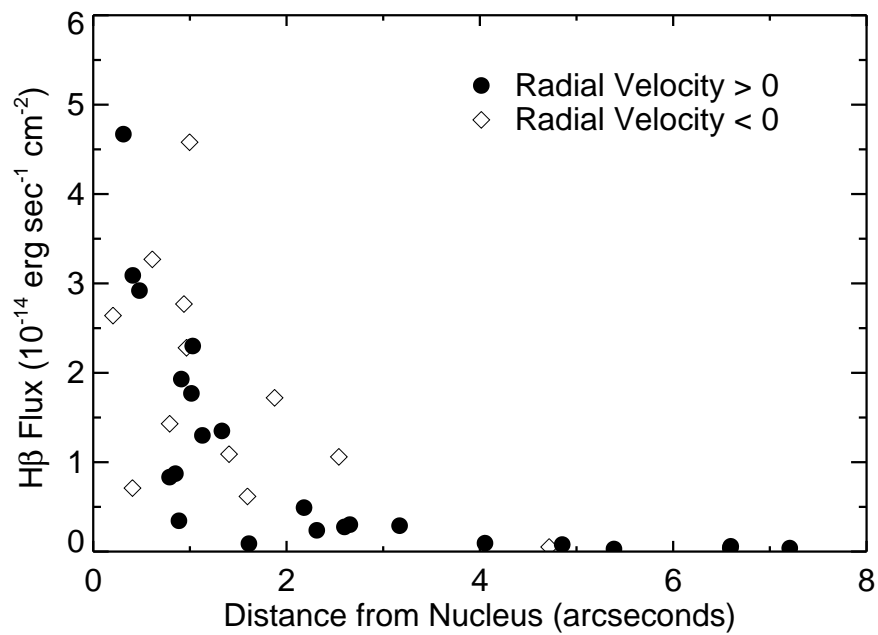
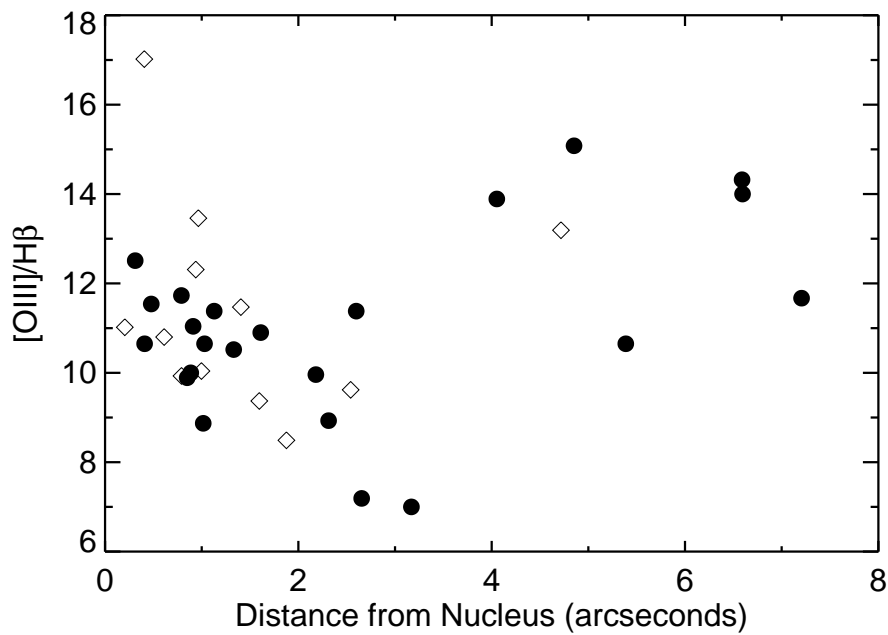
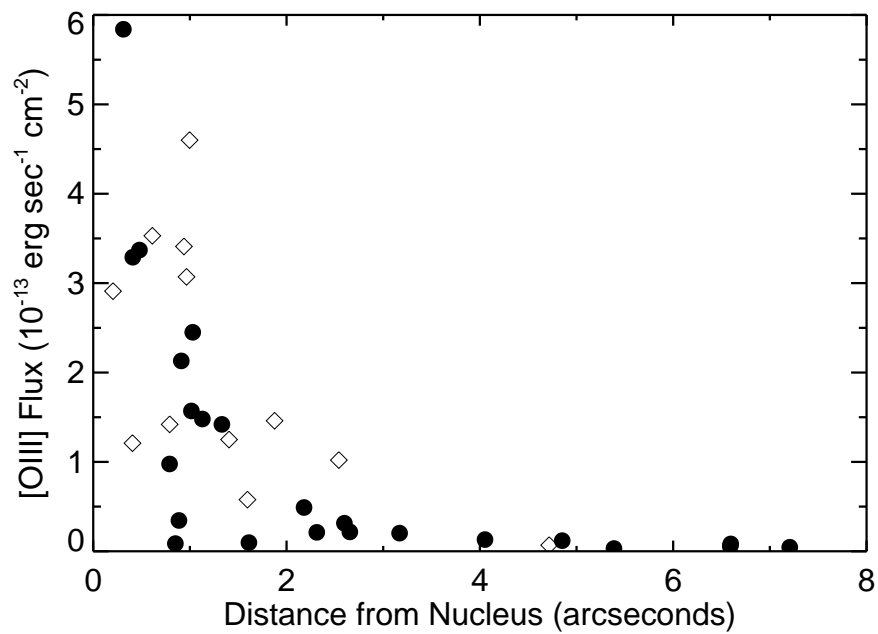
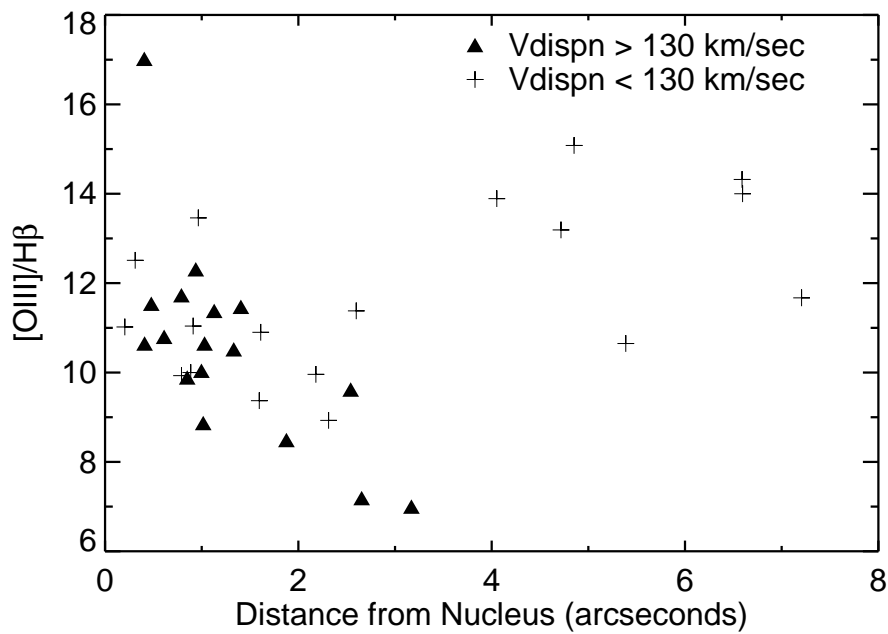
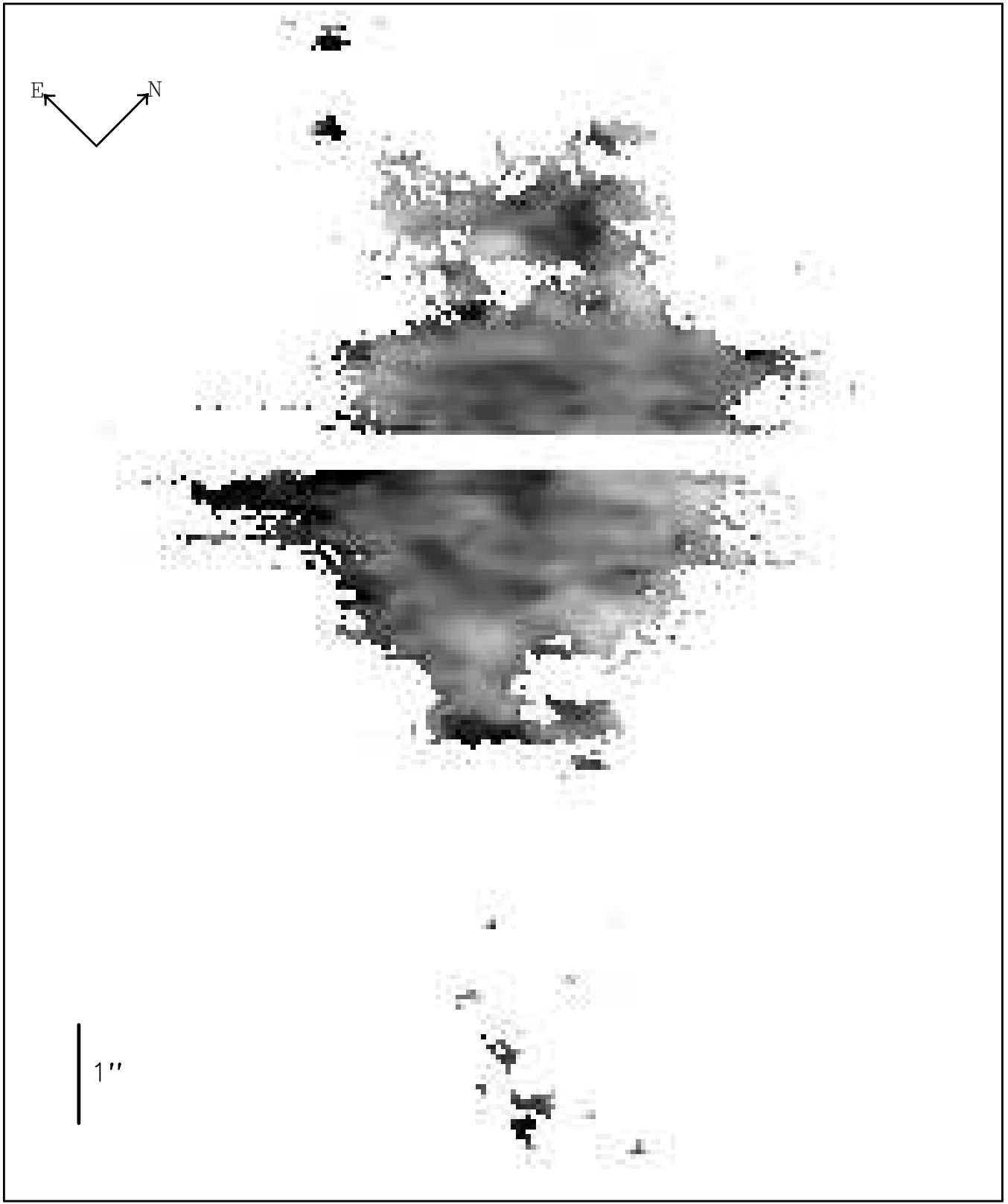
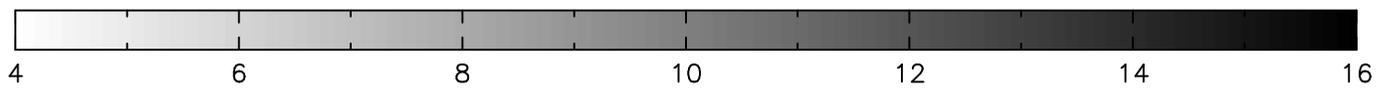
\end{document}